\documentclass[12pt,epsf,amstex]{article}
\usepackage [dvips]{graphicx}
\usepackage{amsmath}
\usepackage{amssymb}
\usepackage{psfig}
\usepackage{epsfig}

\addtocounter{secnumdepth}{1}
\begin{document}
\newcommand{\br}{\bar{r}}
\newcommand{\bbeta}{\bar{\beta}}
\newcommand{\bgamma}{\bar{\gamma}}
\newcommand{\bR}{{\bf{R}}}
\newcommand{\bS}{{\bf{S}}}
\newcommand{\half}{\frac{1}{2}}
\newcommand{\summ}{\sum_{m=1}^n}
\newcommand{\sumqno}{\sum_{q\neq 0}}
\newcommand{\tsum}{\Sigma}
\newcommand{\bsA}{\mathbf{A}}
\newcommand{\bsV}{\mathbf{V}}
\newcommand{\bsE}{\mathbf{E}}
\newcommand{\bsZ}{\hat{\mathbf{Z}}}
\newcommand{\bse}{\mbox{\bf{1}}}
\newcommand{\bspsi}{\hat{\boldsymbol{\psi}}}
\newcommand{\cdottt}{\!\cdot\!}
\newcommand{\deltaR}{\delta\mspace{-1.5mu}R}

\newcommand{\bGamma}{\boldmath$\Gamma$\unboldmath}
\newcommand{\dd}{\mbox{d}}
\newcommand{\ee}{\mbox{e}}
\newcommand{\p}{\partial}

\newcommand{\la}{\langle}
\newcommand{\ra}{\rangle}
\newcommand{\rao}{\rangle\raisebox{-.5ex}{$\!{}_0$}}  
\newcommand{\rae}{\rangle\raisebox{-.5ex}{$\!{}_1$}}
\newcommand{\raG}{\rangle_{_{\!G}}}
\newcommand{\rainr}{\rangle_r^{\rm in}}
\newcommand{\beq}{\begin{equation}}
\newcommand{\eeq}{\end{equation}}
\newcommand{\bea}{\begin{eqnarray}}
\newcommand{\eea}{\end{eqnarray}}
\def\lsim{\:\raisebox{-0.5ex}{$\stackrel{\textstyle<}{\sim}$}\:}
\def\gsim{\:\raisebox{-0.5ex}{$\stackrel{\textstyle>}{\sim}$}\:}

\numberwithin{equation}{section}

\thispagestyle{empty}
\title{{\Large {\bf Asymptotic statistics\\[3mm] 
of the $n$-sided planar Poisson-Voronoi cell.\\[3mm]
I. Exact results\\
\phantom{xxx} }}}
 
\author{{\bf H.\,J. Hilhorst}\\[5mm]
{\small Laboratoire de Physique Th\'eorique,
B\^atiment 210, Universit\'e de Paris-Sud}\\[-1mm]
{\small 91405 Orsay Cedex, France}\\}

\maketitle
\begin{small}
\begin{abstract}
\noindent 

We achieve a detailed understanding 
of the $n$-sided planar Poisson-Voronoi cell
in the limit of large $n$.
Let ${p}_n$ be the probability for a cell
to have $n$ sides.
We construct the asymptotic expansion of 
$\log {p}_n$ up to terms that vanish as $n\to\infty$.
We obtain the statistics of 
the lengths of the perimeter segments 
and of the angles between adjoining segments:
to leading order as $n\to\infty$, and after appropriate scaling, 
these become independent random variables 
whose laws we determine; and to next order in $1/n$
they have nontrivial long range correlations whose expressions we provide.
The $n$-sided cell tends towards a circle
of radius $(n/4\pi\lambda)^{\half}$, where $\lambda$ is the cell density;
hence Lewis' law for the average area $A_n$ of the $n$-sided cell 
behaves as $A_n \simeq cn/\lambda$ with $c=\frac{1}{4}$.
For $n\to\infty$
the cell perimeter, expressed as a function $R(\phi)$ 
of the polar angle $\phi$,
satisfies $\dd^2 R/\dd\phi^2 = F(\phi)$, 
where $F$ is known Gaussian noise; 
we deduce from it the probability law for the perimeter's
long wavelength deviations from circularity. 
Many other quantities
related to the asymptotic cell shape become 
accessible to calculation.\\

\end{abstract}
\end{small}
\vspace{5mm}

\noindent LPT -- ORSAY 05/42\\
{\small $^1$Laboratoire associ\'e au Centre National de la
Recherche Scientifique - UMR 8627}
\newpage


\section{Introduction} 
\label{secintroduction}
\vspace{5mm}

Experimental data on cellular structures
are often compared to the properties
of a {\it Voronoi diagram} \cite{Voronoi08}, sometimes also called a 
Voronoi {\it tessellation.}  
This well-defined mathematical object 
is a partitioning of space into convex cells
constructed around `point particles' (also called `centers,' `seeds,'
or `germs') 
in such a way that each point of space
is in the cell of the particle to which it is closest. 
When the distribution of the point particles is {\it uniformly random}, 
the result is said to be a {\it Poisson}-Voronoi diagram;
it is to  naturally occurring tessellations 
what the ideal gas is to real gases.
\vspace{3mm}

In this work we will be interested in planar cellular structures. 
An example analyzed
more than seventy years ago by Lewis \cite{Lewis}
is the epidermal epithelium of cucumber,
which is composed of many flat, tile-like cells.
For other applications of Voronoi diagrams outside of physics we refer
to the encyclopedic work by Okabe {\it et al.} \cite{Okabeetal00}.

Within physics, two-dimensional cellular structures
appear in many different experimental situations, of which we
mention only a few.
Cerisier {\it et al.} \cite{CRR96} studied the cellular patterns in
surface-tension driven B\'enard convection.
Elias {\it et al.} \cite{EFBCG97} studied two-dimensional magnetic liquid
froth. 
Sire and Seul \cite{SireSeul95}
analyzed droplet patterns in late-stage coarsening of a 
two-dimensional binary mixture. 
Moriarty {\it et al.} \cite{MTB02} analyzed nanostructured cellular
layers by means of the Voronoi construction.
Earnshaw and Robinson \cite{EarnshawRobinson94,EarnshawRobinson95} 
performed a Voronoi cell analysis of the structure of two-dimensional
colloidal aggregates. Much
important experimental work on tessellations comes from two-dimensional 
soap froths, as studied experimentally
by Glazier {\it et al.} \cite{GGS87} and by Stavans and Glazier
\cite{StavansGlazier89}.
In simulations of the two-dimensional
solid-liquid transition (recent examples are
Zahn {\it et al.} \cite{ZLM99} and Quinn and Goree \cite{QuinnGoree01}),  
Voronoi cells constructed
around the particle positions define the
nearest-neighbor relations between atoms, which in turn allow the
identification of lattice defects. 
Richard {\it et al.} \cite{ROTG99} used the Voronoi construction for the
geometrical characterization of hard sphere systems.

Evidently, there are many different reasons 
why none of the above systems should be
described exactly by a Poisson-Voronoi diagram.
In dense fluids the particle positions are correlated
and the resulting Voronoi diagrams are not of the Poisson type.
Both soap froths and binary mixtures coarsen with time,
and the experimentally determined properties of these 
cellular structures usually refer to a stationary
scale invariant distribution (sometimes called the
`equilibrium distribution') observed at long
times. Clearly such systems are best analyzed, in principle, 
by a dynamical theory such as developed, {\it e.g.,} by Marder \cite{Marder87},
by Stavans {\it et al.} \cite{SDM91},
and by Flyvbjerg \cite{Flyvbjerg93}.
Nevertheless, comparison of dynamical experimental data to the
static Poisson-Voronoi
diagrams may give a clue as to what elements should enter
into a correct dynamical theory.

Discrepancies between the experimental data and the predictions of a
Poisson-Voronoi diagram are often
interpreted as the consequence of some
physical, biological (see {\it e.g.} \cite{JeuneBarabe98}), 
or other action that one would like to identify.
It is therefore of importance that the properties of Poisson-Voronoi 
diagrams be understood as well as possible. 

In field theory `pure' Voronoi diagrams play a role since
attempts were made \cite{Christetal82,ID89}
to construct a theory on a lattice of randomly located sites.
The Voronoi construction then again determines 
the nearest neighbor pairs among the sites. 
Such a random lattice has the
advantage of being statistically invariant under
arbitrary translations and rotations while still preserving
a short-distance cutoff. 
Field theory provides
still another reason for interest in Voronoi diagrams:
Godr\`eche {\it et al.} \cite{GKY92} showed that
the ensemble of cells with trivalent vertices in a plane
is related to the problem of counting planar Feynman diagrams 
(`Feynman foam') of a field theory with cubic interaction.


\subsection{Statistics of planar Poisson-Voronoi cells}
\label{secplanar}

Planar Poisson-Voronoi diagrams are characterized
by their statistical properties.
The first and foremost of these, whether in experimental studies,
in theoretical work , or in simulations,
is the probability $p_n$ that a cell have $n$ sides.
Others include the average number $\la n \ra$ of cell sides; 
the average area $A_n$ of an $n$-sided cell;
the statistics of the cell perimeter and of its angles;
and correlations between neighboring or nearby cells. 
The topology of the plane imposes that $\la n\ra=6$.
One may readily obtain several of the other statistical
properties analytically by expressing them
as integrals on the positions of the point particles
\cite{Meijering53,Okabeetal00}.
The calculation of the fraction $p_n$, however,
has still defied solution \cite{ID89,Okabeetal00}. 
The distribution $p_n$ peaks at $n=6$ and falls of rapidly for large $n$.
For $n=3$ Hayen and Quine \cite{HayenQuine00}
have expressed $p_3$ as a fivefold integral which they 
evaluated numerically to seven decimal places. 
For $n=4,5,\ldots$ the values for $p_n$
obtained from Monte Carlo simulations have been tabulated. 
All Monte Carlo studies cited in reference \cite{Okabeetal00}
(see {\it e.g.} \cite{QuineWatson84,LeCaerHo90,KumarKurtz93}) 
were restricted to $n\lsim 14$,
which is the largest number of sides observed in simulations by conventional
algorithms: one has $p_{14}\approx 10^{-6}$.

As early as 1984 Drouffe and Itzykson 
\cite{DI84}, by a then novel method,
estimated the probability ${p}_n$ for $n$ up to $50$ 
by Monte Carlo simulations.
This work is still today the main
reference for the large $n$ behavior. 
The presently known values of $p_n$ have their uncertainty typically in the
third digit for $4 \leq n \lesssim 10$ and in the second digit
for $10 \lesssim n \lesssim 16$.
Very  large cells with $n$ values higher than 60
were observed by Lauritsen {\it et al.}
\cite{LMH93} in simulations of a
`Voronoi cell Hamiltonian' which, although of uncertain physical relevance,
is interesting because it favors the appearance
many-sided cells; this procedure 
does not, however, provide estimates for $p_n$.
\vspace{3mm}

In view of its importance as a system of reference, we study in this
work the planar Poisson-Voronoi diagram. 
We obtain a great many new analytic
results for $n$-sided cells in 
an asymptotic expansion for large $n$.

Expansions around idealized models are common in physics and as such
our approach needs no specific justification. The fact that
$n=\infty$ is not physically accessible in any realization 
known of today, is of no concern, 
since the expansion goes into the physical domain $n<\infty$.
After the work is done, we will point out the relevance 
of our results for finite $n$. In particular,
the vast body of experimental and numerical
results has led to two empirical laws that are
satisfied to a good extent by many systems: Lewis' law \cite{Lewis} and 
Aboav's law \cite{Aboav70}. 
The first one will be discussed at the end of this paper; 
the second one will be investigated separately \cite{HJHaboav05}
as an extension of our present work.

An announcement of part of our results, including a
concise indication of their derivation, 
has appeared earlier \cite{HJHletter05}.
Here we present the full calculation 
together with many more results. 
This is Part I of a series of two papers. 
It presents our analytic
work and the results that derive immediately from it.
In Part II \cite{HJHpartII05}
we will develop arguments which,
with the benefit of hindsight,
will help to understand our results heuristically
and to employ them in other applications.
\vspace{2mm}


\subsection{Method}
\label{secmethod}

We consider a square $L\times L$ subdomain of the plane $\mathbb{R}^2$.
In this domain ${N}$ point particles 
are placed at random positions, 
chosen independently
and with  uniform probability. 
We consider the limit ${N},{L}\to\infty$    
with the particle density ${N/L^2}=\lambda$ fixed.
In mathematical terminology this is the {\it Poisson 
point process of intensity $\lambda$} in the plane.
The density $\lambda$ may be trivially scaled away.
The Voronoi cells of this point particle system 
are necessarily convex polygons; their
number of sides $n$ may be any integer $\geq 3$.

We ask about the statistical
properties of the Voronoi cell
of an arbitrarily selected particle,
whose position is taken as the origin of the coordinate system.
Let the other particles, indexed by a label $a$,
have position vectors $2\bR_a$.
The perpendicular bisectors of these vectors pass through the 
{\it midpoints} $\bR_a$,
and the line segments that constitute the cell sides lie on these bisectors.
Out of the full set of bisectors only a finite number $\nu$
that pass sufficiently close to the origin
will actually
contribute a segment to the perimeter of the cell containing
the origin. We denote these `contributing' midpoints as
$\bR_1,\bR_2,\ldots,\bR_\nu$.
They determine the positions of the cell vertices, 
which we indicate by $\bS_1,\bS_2,\ldots,\bS_\nu$.
The angles between two successive midpoint vectors $\bR_{m-1}$ and $\bR_{m}$ 
[vertex vectors $\bS_{m-1}$ and $\bS_{m}$] 
will be denoted as $\xi_m$ [as $\eta_m$].
The probability $p_n$ that we have $\nu=n$ for a given $n$,
{\it i.e.,} that the selected Voronoi cell
have exactly $n$ sides, can be expressed as a $2n$-dimensional integral on
$\bR_1,\bR_2,\ldots,\bR_n$.

We evaluate this integral 
for asymptotically large $n$ by a
multidimensional steepest descent method.
Although the principle of this approach is simple, its
execution is beset by numerous interdependent complications.
The necessity to deal with these is the reason for the length of this paper.
We were initially guided by the idea of describing the perimeter 
in terms of a Markov process
as a function of the polar angle; the Markov aspect
of our description will, however, stay in the background
and we will not seek to elaborate on it. 
\vspace{2mm}

We write
$p_n={\rm Tr\,}\ee^{-{\mathbb H}},$ where ${\mathbb H}$ is a `Hamiltonian'
of purely geometrical origin
and $\,{\rm Tr}\,$ stands for the integrations.
For a suitably chosen `free' or `noninteracting' Hamiltonian 
${\mathbb H}_{\,0}$ we set 
${\mathbb H}={\mathbb H}_{\,0}+{\mathbb V}$ and write
\beq
p_n = \big({\rm Tr\,} \ee^{-{\mathbb H}_{\,0}}\big)\,
\langle\ee^{-\mathbb{V}}\rao\,,
\label{defordersintro}
\eeq
where $\la\ldots\rao$ denotes the average with respect to 
$\ee^{-{\mathbb H}_{\,0}}$.
We will refer to the two factors, ${\rm Tr\,}\ee^{-{\mathbb H}_{\,0}}$
and $\langle\ee^{-\mathbb{V}}\rao$, as the 
`noninteracting' (or `free') and the `interacting' problem,
respectively.
For clarity of presentation, rather than for technical reasons,
we study them separately.

In the course of our analysis
the appropriate ${\mathbb H}_{\,0}$ appears to be a sum of noninteracting
terms in the angles $\xi_m$ and $\eta_m$:
for $n\to\infty$ these variables decouple.
The evaluation of  
${\rm Tr\,}\ee^{-{\mathbb H}_{\,0}}$
amounts to a $2n$-dimensional steepest descent problem in
the positive orthant ({\it i.e.} the subspace with all $\xi_m,\eta_m>0$),
with contributions concentrated in a boundary layer of width
of order $n^{-1}$.
The answer can be obtained exactly up to corrections that vanish
exponentially as $n\to\infty$. 
 
The result obtained for ${\rm Tr\,}\ee^{-{\mathbb H}_{\,0}}$
is meaningful only if we are able to show that 
the calculation of $\langle\ee^{-\mathbb{V}}\rao$
actually leads to a controlled perturbation series in
negative powers of $n$.
The expression for the `interaction' ${\mathbb V}$ 
is so opaque that {\it a priori\,} it is not clear
whether it lends itself to any kind of
large $n$ expansion, let alone how it will contribute 
to the $n$ dependence of the final result.

We overcome this problem by considering the perimeter segments as 
a set of vectors $\bS_{m}-\bS_{m-1}$ which,
by analogy to the theory of elasticity,
we expand about their values in 
a regular $n$-sided polygon. 
An analysis of considerable difficulty shows that 
we may write ${\mathbb V}={\mathbb V}_1+n^{-\half}{\mathbb V}_2$, where
${\mathbb V}_1$ is a quadratic form in terms of the `elastic deviations',
and ${n^{-\half}\mathbb V}_2$ collects together 
a great number of higher order terms. The maximum of 
$\,\ee^{-{\mathbb V}_1}\,$ is located 
in the positive orthant, {\it inside} the boundary layer identified in
the free problem. The `elastic' interaction ${\mathbb V}_1$
has a coupling constant proportional to $n^{-1}$, 
but is long-ranged along the perimeter
in the same way as a one-dimensional
Coulomb interaction.
There is a full separation of spatial scales: 
${\mathbb H}_{\,0}$ is a Hamiltonian of
independent `microscopic' variables; and ${\mathbb V}_1$
has its weight concentrated on
an effectively finite number of wavelengths at the scale of the cell.

The integrations in the interacting problem involve the confluence
of the boundary layer maximum and the maximum of $\,\ee^{-{\mathbb V}_1}\,$. 
They can be dealt with via a Hubbard-Stratonovich transformation.
Due to the long-rangedness of the interaction and the weakness
of the coupling, an expansion in negative powers of $n$ becomes possible.
In the end we find
that the leading order term in the interacting problem 
is equal to a finite constant:
$\la\ee^{-\mathbb{V}}\rao\simeq C$ as $n\to\infty$.
\vspace{2mm}

Various technical obstacles surge in the course of our analysis. 
To conclude this subsection, we mention three of them.

First,
in previous approaches \cite{DI84,Calka03a,Calka03b}
the domain of integration in phase space
was always defined in terms of a set of $n$ inequalities.
Here we choose variables of integration such that it becomes fully explicit.
After suitable transformation
the integration bears on only one radial variable
and $2n-1$ angles. 
The radial integration can be carried out exactly
and the remaining angular integrations constitute the true problem.
There is no single set of angular variables in terms of which the integrand
takes a simple form.

Secondly, 
the angular variables satisfy sum rules stemming from various different
origins. It is indispensable to strictly keep track of each of them. 
To that end
we denote them as `geometrical sum rules,' a `gauge condition,' and a
`no-spiral' constraint; their meanings will be explained as they are
encountered. 

Thirdly, 
the asymptotic expansion for $n\to\infty$
requires, as usually, that we postulate at the outset the
appropriate scaling with $n$ for the variables of integration.
But since $n$ is also the typical number of terms 
in many summations that occur, 
our scaling analysis has to deal with these sums, too.
Some remarkable cases of canceling leading orders
need particular attention.
\vspace{2mm}

Throughout this work we use only those methods of
mathematical physics that ordinarily lead to exact results.


\subsection{Results}
\label{secresults}

We obtain a detailed understanding of the statistical properties 
of the $n$-sided Voronoi cell when $n$ is large.
The main qualitative results are summarized as follows.

(i) 
The probabilty ${p}_n$
for a Voronoi cell to have $n$ sides behaves
asymptotically as
\beq
{p}_n = \frac{C}{4\pi^2}\,\frac{(8\pi^2)^n}{(2n)!} 
\,\big[1+{\cal O}(n^{-\frac{1}{2}})\big], \qquad n\to\infty,
\label{resultpn}
\eeq
where
\begin{align}
C &= \prod_{q=1}^\infty\,\Big( 1-\frac{1}{q^2}+\frac{4}{q^4} 
\Big)^{-1}
\nonumber\\[1mm]
  &= 0.344\,347...
\label{resultC}
\end{align}
and where ${\cal O}(n^d)$ denotes a term which for $n\to\infty$ is to
leading order proportional to $n^d$.
Alternatively this may be written as
\beq
\log p_n = -2n\log n + n\log (2\pi^2\ee^2) 
- \tfrac{1}{2}\log \big( 2^6\pi^5 C^{-2}n \big) 
+ {\cal O}(n^{-\half}).
\label{resultlogpn}
\eeq
In (\ref{resultpn}) the factors $(4\pi^2)^{-1}(8\pi^2)^n/(2n)!$ 
and $C$ originate from the free and the interacting problem,
respectively, {\it i.e.,} from the factors
${\rm Tr\,}\ee^{-{\mathbb H}_{\,0}}$
and $\langle\ee^{-\mathbb{V}}\rao$
in (\ref{defordersintro}).
The constant $C$ in carries a great deal of meaning.
Its factor with index $q$ stems 
from the `elastic' modes with wavelength
$2\pi/q$. Hence the contributions to $C$ are seen to come
essentially from the large scale, long
wavelength modes.
With the decrease of scale they damp out 
as $\sim 1/q^2$.
\vspace{2mm}

The remaining results, (ii)-(vii) below,
arise as byproducts of the calculation of $p_n$.
They have, nevertheless, an independent interest
equal to that of $p_n$, if not greater.

\begin{figure}
\begin{center}
\scalebox{.60}
{\includegraphics{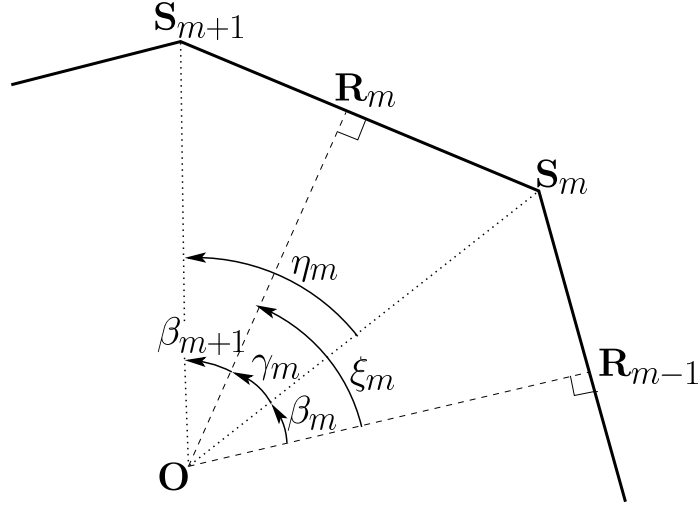}}
\end{center}
\caption{{\small Heavy line: the perimeter of the Voronoi cell of the
particle in the origin. Dashed and dotted lines connect the origin to the
midpoints and vertices, respectively. Right angles have been marked.
The figure defines the
angles $\xi_m$ and $\eta_m$ discussed in this section as well as 
$\beta_m$ and $\gamma_m$ to be introduced in section 2.}}
\label{fig1}
\end{figure}

(ii) 
As shown in figure \ref{fig1}, we write
$\xi_m$ for the angle by which the perimeter
turns at the $m$th vertex (it is also the angle between two successive
midpoint vectors); and 
$\eta_m$ for the angle by which it
advances as it passes from the $m$th to the $(m+1)$th vertex.
The $\xi_m$ are positive and add up to $2\pi$, and the $\eta_m$ similarly. 
Because of these sum rules 
they have a weak `background' correlation which
in the limit $n\to\infty$  vanishes as $\sim n^{-1}$ 
(here and henceforth, the sign $\sim$ denotes asymptotic proportionality).
In that limit the $\xi_m$ and
$\eta_{m}$ become independent random variables with
the $\xi_m$ identically distributed according to the law 
\beq
u(\xi)=\frac{n^2\xi}{\pi^2}\,\exp\big(\!-\frac{n\xi}{\pi} \big)
\label{resuxi}
\eeq  
and the $\eta_m$ according to
\beq
v(\eta)=\frac{n}{2\pi}\exp\big(\!-\frac{n\eta}{2\pi} \big).
\label{resveta}
\eeq
These distributions have the averages
$\bar{\xi}=\bar{\eta}=2\pi n^{-1}$, as had to be the case.
The background correlations have contributions not only from the sum rules,
but also from the interaction ${\mathbb V}$, which renders them nontrivial.
We find the expressions for these correlations in sections 
\ref{secV} and \ref{secdeviations} but will 
not state the results here. 

(iii) 
Let the mean value of the midpoint distances  
$R_1,R_2,$ $\ldots, R_n$ 
be denoted by $R_{\rm av} = n^{-1}\sum_{m=1}^nR_m$ and let
\beq
R_{\rm av} = R_c + \deltaR_{\rm av}, \qquad
R_c=\Big( \frac{n}{4\pi\lambda} \Big)^\half.
\label{defRc}
\eeq
In the limit $n\to\infty$ we have $\la R_{\rm av}\ra=R_c$ and
$\deltaR_{\rm av}$ has the Gaussian distribution 
\beq
{\sf p}(\deltaR_{\rm av})= (8\lambda)^\half
\exp\big( \!-8\pi\lambda\, \deltaR_{\rm av}^2 \big).
\label{defpRav}
\eeq
{\it i.e.,} its width is independent of $n$.

(iv)
The fluctuations of the $R_m$ around their average $R_{\rm av}$
have the $n$ independent variance
\beq 
\langle (R_m-R_{\rm av})^2\rangle \simeq \frac{3C_{\rm av}}{4\pi\lambda},
\qquad n\to\infty,
\label{varRm}
\eeq
with $C_{\rm av}=\sum_{q=1}^\infty q^{-4}(1-q^{-2}+4q^{-4})^{-1}\
=0.333\,418...$.
Here and henceforth $\la\ldots\ra$ indicates an average over
all Voronoi diagrams and $\simeq$ stands for asymptotic equality.
Equations (\ref{defRc}) and (\ref{varRm}) together
tell us that the $R_m$ all scale as $n^{\half}$ 
and that the line segments linking
the midpoints $\bR_m$
run mostly inside an imaginary tube of width $\sim n^0$
along a circle of radius $R_c=(4\pi\lambda)^{-\half}n^{\half}$. 
The cell {\it perimeter\,} is the polygon 
having not the $\bR_m$, but the
$\bS_m$ as vertices; however, we show that the distance between the two
polygons vanishes as $\sim n^{-\half}$.
Therefore the cell perimeter
approaches a circle as $n\to\infty$.

This approach to circularity when the {\it number of sides\,}
tends to infinity is
distinct from other approaches to circularity,
such as occur for cells 
whose {\it area\,} tends to infinity
(see Hug, Reitzner, and Schneider \cite{Hugetal04}),
or whose {\it inradius} ({\it i.e.,} the radius of the largest inscribed
circle centered at the origin)
tends to infinity (see Calka \cite{Calka02};
Calka and Schreiber \cite{CalkaSchreiber05}). 
In Part II \cite{HJHpartII05} 
we compare the two types of limits in greater detail.
 
(v) 
Let $2\pi mn^{-1}=\phi$ and $R_m=R(\phi)$.
In the limit $n\to\infty$ the variable $\phi$ becomes
continuous and may be identified, up to corrections that vanish as
$n^{-\half}$, with the polar angle of the perimeter.
We set 
\beq
R(\phi) = R_c + \deltaR(\phi).
\label{defdeltaR}
\eeq
and refer to $\deltaR(\phi)$ as the `excess midpoint distance.' 
On scales large compared to the angular scale $2\pi n^{-1}$
of the individual perimeter segments the following holds.
The effect of the Poisson point process on the perimeter $R(\phi)$
may be represented by a stochastic
second order differential equation for $\deltaR(\phi)$,
\beq
\frac{\dd^2 \deltaR}{\dd\phi^2}=F(\phi), \qquad
0<\phi<2\pi,
\label{oderphi}
\eeq
in which $F(\phi)$ is
zero average Gaussian noise of autocorrelation
\beq
\la F(\phi)F(\phi') \ra = 12\lambda
\Big[ \delta(\phi-\phi') \,-\, \frac{1-\Gamma(\phi-\phi')}{2\pi} \Big],
\label{corrF}
\eeq
where $\Gamma(\phi)$ is a $2\pi$-periodic function
whose integral on $[0,2\pi]$ vanishes.
The solution of (\ref{oderphi}) is
subject to a periodicity condition 
and an integral condition, {\it viz.}
\beq
\deltaR(0)=\deltaR(2\pi),
\qquad
\frac{1}{2\pi}\int_0^{2\pi}\!\!\dd\phi\,\deltaR(\phi)=\deltaR_{\rm av}, 
\label{integralcond}
\eeq
respectively, with
$\deltaR_{\rm av}$ distributed according to (\ref{defpRav}). 
Hence in the $n\to\infty$ limit
$\deltaR(\phi)$ is governed by Gaussian statistics.
The correlation function (\ref{corrF}) determines the
probability law 
of the noise $F(\phi)$ and hence, by equations (\ref{oderphi}),
(\ref{integralcond}), 
and (\ref{defpRav}), of $\deltaR(\phi)$. The explicit result
is given in section \ref{secRphi}.

(vi) 
According to Lewis' empirical law \cite{Lewis,Okabeetal00}
the average area $A_n$ of 
an $n$-sided cell grows linearly with $n$.
For Poisson-Voronoi cells, in spite of attempts, this 
linearity with $n$ has not so far been proved
\cite{RivierLissowski82,Rivier85}. 
Now the results (iii) and (iv) above imply directly that
\beq
A_n \simeq \pi R_c^2 
= \frac{n}{4\lambda}, \qquad n\to\infty.
\label{Anaspt}
\eeq
Hence we have derived here from first principles that Lewis' law 
holds in the asymptotic regime and has coefficient $\frac{1}{4}$.


\section{Writing ${p}_n$ as a $2n$-fold angular integral}
\label{secangularvariables}

In this section we start from the basic $2n$-fold integral for $p_n$ and
subject it to a sequence of transformations that will cast it in a form
amenable to an asymptotic large $n$ expansion. 


\subsection {Basic integral for $p_n$}

An $n$-sided cell is completely defined by 
$n$ midpoint positions $\bR_1,\bR_2,\ldots,\bR_n$ (see figure \ref{fig1}).
These are uniformly distributed with density $4\lambda$,
and $p_n$ may be written directly as a $2n$-fold integral on their
coordinates \cite{DI84,ID89,Calka03a,Calka03b},
\begin{equation}
{p}_n=\frac{(4\lambda)^n}{n!}
\int \dd\bR_1 \ldots \dd\bR_n\,\,
\chi(\bR_1,\ldots,\bR_n)\,\, 
\ee^{-4\lambda{\cal A}(\bR_1,\ldots,\bR_n)},
\label{exprR2pn}
\end{equation}
where the functions $\chi$ and ${\cal A}$
to be defined shortly enforce two
conditions that must be satisfied if the $n$ points are to define a
valid Voronoi cell. In order to state these conditions 
we use the polar coordinate representation $\bR_m=(R_m,\Phi_m)$.
In the integrand of (\ref{exprR2pn}) we may
set one of the angles, say $\Phi_n$, to zero if we compensate by an extra
factor $2\pi$, and we may order the remaining angles according to 
\beq
\Phi_n = 0<\Phi_1 < \Phi_2 < \ldots < \Phi_{n-1} < 2\pi
\label{orderOmegam}
\eeq
if a compensating factor $(n-1)!$ is introduced.
We will denote the $n$ angular differences $\xi_m$ 
between successive midpoint vectors as
\bea
&&\xi_m = \Phi_m - \Phi_{m-1}\,, \qquad m=1,\ldots,n-1,\nonumber\\
&&\xi_n = 2\pi - \Phi_{n-1}\,.
\label{defxim}
\eea
The geometry constrains the $\xi_m$ to the
interval $(0,\pi)$ and they must satisfy the geometrical sum rule 
\begin{equation} 
\sum_{m=1}^{n}\xi_m=2\pi.
\label{sumrulexi}
\end{equation}
We find it convenient henceforth to scale the distances such that 
\beq
4\lambda=1
\label{scalinglambda}
\eeq
during the calculations,
and restore this factor in the final results.
Symmetrizing in the angular difference variables by introducing a
Dirac delta function we obtain from (\ref{exprR2pn})
\beq
{p}_n=\frac{2\pi}{n}
\int_0^\infty\!\!R_1\dd R_1\ldots R_n\dd R_n
\int_0^\pi\!\!\dd\xi_1\ldots\dd\xi_n\,\, 
\delta(\xi_1+\ldots+\xi_n-2\pi)\,
\chi\,\ee^{-{\cal A}}.
\label{exprRxipn}
\eeq
We will also need the positions $\bS_m=(S_m,\Psi_m)$ of the cell  vertices
(see figure \ref{fig1}). These are auxiliary variables in the sense that
they can be expressed in terms of the $\xi_\ell$ and $R_\ell$.
We are now ready to state the two conditions.

{\it Condition 1.}
The first condition is that for $a\neq 1,\ldots,n$ 
none of the bisectors passing through the midpoints $\bR_a$ 
have an intersection with the Voronoi cell.
This condition is fulfilled if these midpoints stay outside of a 
region of the plane equal to the union
of $n$ disks centered at $\frac{1}{2}\bS_m$ and of radius
$\frac{1}{2}S_m$ \cite{footnote1}. 
If this region has area ${\cal A}$, then
the probability for it to be empty of midpoints
(other than the $n$ that determine the cell perimeter)
is equal to\, $\exp(-{\cal A})$\, times its area. 
An explicit expression for ${\cal A}$ in terms of the $R_m$ and $\xi_m$ may be
found by straightforward geometry (see Calka \cite{Calka03a,Calka03b}). 

{\it Condition 2.}
The second condition defines the domain of integration. It says that
each of the $n$ midpoints $\bR_m$ should have 
its corresponding
perpendicular bisector contribute a nonzero segment to the perimeter. 
This condition was shown by Calka \cite{Calka03a,Calka03b} to read
explicitly
\begin{equation}
R_{m-1}\sin\xi_{m+1}+R_{m+1}\sin\xi_{m}
> R_m\sin(\xi_m+\xi_{m+1}), \quad m=1,\ldots,n,
\label{condition2}
\end{equation}
with the conventions $R_0=R_n$,\, 
$R_{1}=R_{n+1}$, and $\xi_{1}=\xi_{n+1}$.
The function $\chi$ in (\ref{exprR2pn}) and (\ref{exprRxipn}) 
is the indicator of the domain in phase space where 
(\ref{condition2}) is satisfied.

This fully defines the integral (\ref{exprRxipn}) for $p_n$.
In being a sum on $n$ equivalent interacting objects, 
{\it viz.} the midpoint vectors $\bR_m$, 
this integral is analogous to a partition
function in statistical mechanics. But although interesting,
this observation provides no direct key as to how to go about its
evaluation.


\subsection{Angular variables}
\label{sectwosets}

There is no single set of variables in which the
calculations of this work take a simple form.
We will, throughout, employ
simultaneously different sets of angular variables,
to be introduced in this subsection.

\begin{figure}
\begin{center}
\scalebox{.60}
{\includegraphics{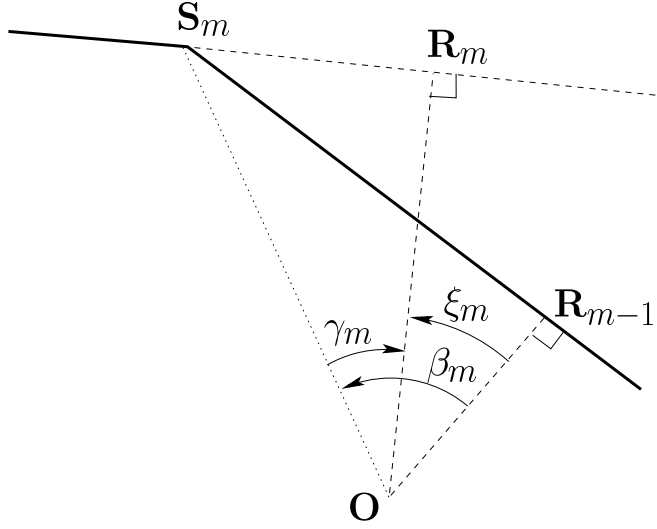}}
\end{center}
\caption{{\small A configuration of midpoints $\bR_m$ and $\bR_{m+1}$
leading to a negative angle $\gamma_m$. 
The relation $\beta_m+\gamma_m=\xi_m$ remains valid.
The negativity of $\gamma_m$ is accompanied by a
counterclockwise outward spiraling of the perimeter. 
The legend is the same as for figure 1.}}
\label{fig2}
\end{figure}

The polar angles $\Psi_m$ of the vertex vectors $\bS_m$ satisfy
an ordering analogous to (\ref{orderOmegam}),
\beq
0<\Psi_1<\Psi_2<\ldots<\Psi_n<2\pi.
\label{orderPsim}
\eeq
There is, however, no particular ordering relation between the
two sets $\{\Psi_m\,|\,m=1,\ldots,n\}$ and $\{\Phi_m\,|\,m=1,\ldots,n\}$. 
We will not use the $\{\Psi_m\,|\,m=1,\ldots,n\}$ any more, but instead, 
by analogy to equation (\ref{defxim}), introduce the angular differences 
\bea
\eta_m &=& \Psi_m-\Psi_{m-1}\,, \qquad m=2,\ldots,n,\nonumber\\
\eta_1 &=& \Psi_1-\Psi_n + 2\pi,
\label{defetam}
\eea
which are constrained to $(0,\pi)$ and satisfy the geometrical sum rule 
\begin{equation} 
\sum_{m=1}^n\eta_m=2\pi.
\label{sumruleeta}
\end{equation}
Next, as shown in figure \ref{fig1},
the set $\{\beta_m,\gamma_m\,|\,m=1,\ldots,n\}$ will represent the angles
between 
a vertex vector and a midpoint vector.
They are defined in terms of the $\Phi_m$ and $\Psi_m$ by
\bea
\beta_m &=& \Psi_m-\Phi_{m-1}\,, \qquad m=2,\ldots,n, \nonumber\\
\beta_1 &=& \Psi_1\,, \nonumber\\
\gamma_m &=& \Phi_m-\Psi_m\,, \qquad m=1,\ldots,n.
\label{defbetagammam}
\eea
The $\beta_m$ and $\gamma_m$ are geometrically constrained
to the interval $(-\frac{\pi}{2},\frac{\pi}{2})$, and hence,
contrary to the $\xi_m$ and $\eta_m$, may be negative.
An example of a negative $\gamma_m$ is shown in figure \ref{fig2}.
Whereas this potential negativity may at this point not seem worthy of much
attention, it will be seen in section \ref{secmorescaling} to have important 
consequences for the scaling of these angles with $n$.
Because of the geometrical sum rule
\beq
\sum_{m=1}^n(\beta_m + \gamma_m) = 2\pi,
\label{sumrulebetagamma}
\eeq
the set $\{\beta_m,\gamma_m\,|\,m=1,\ldots,n\}$ contains one redundant angle.
Hence, leaving out $\gamma_n$, the set of $2n-1$ variables 
\beq
\{\beta_1,\gamma_1,\ldots,\beta_{n-1},\gamma_{n-1},\beta_n\}
\label{defsetbetagamma}
\eeq
is complete in the sense that, given $\Phi_0=0$, all
the other angles $\Phi_m$ and $\Psi_m$ are easily
expressed in terms of the angles of the set (\ref{defsetbetagamma}).

The set $\{\xi_m,\eta_m\,|\,m=1,\ldots,n\}$, 
due to the sum rules (\ref{sumrulexi}) and (\ref{sumruleeta}),
contains two redundant variables and
is not complete:
it specifies only the angles of the $\bR_m$ between themselves and of
the $\bS_m$ between themselves, but not the relative angle between the
two vector systems.
To fix their relative orientation, one other angle is
needed, for which we will take $\beta_1$. 
A second complete set of $2n-1$ variables is therefore
\beq
\{\xi_1,\eta_1,\ldots,\xi_{n-1},\eta_{n-1},\beta_1\}.
\label{defsetxieta}
\eeq 
The sets (\ref{defsetbetagamma}) and (\ref{defsetxieta}) 
have $\beta_1$ in common. The 
transformation relation between their $2n-2$ remaining variables is
\beq
\xi_m=\beta_m+\gamma_m, \qquad \eta_m=\gamma_m+\beta_{m+1}\,,
\qquad m=1,\ldots,n-1,
\label{relxietabetagamma} 
\eeq
and its nonlocal inverse
\bea
\beta_m &=& \phantom{-}\beta_1 - 
\sum_{\ell=1}^{m-1}(\xi_{\ell}-\eta_\ell), 
\qquad m=2,\ldots,n, \nonumber\\
\gamma_m &=& -\beta_1 +
\sum_{\ell=1}^{m-1}(\xi_{\ell}-\eta_{\ell}) + \xi_m\,, 
\qquad m=1,\ldots,n-1.
\label{inversebgxy}
\eea
where by definition $\sum_{\ell=1}^0=0$ for any summand.
It is easy to verify that this transformation 
has unit Jacobian. 
Relations (\ref{relxietabetagamma}) are valid also for $m=n$
with the convention that $\beta_{n+1}=\beta_1$; and
relations (\ref{inversebgxy}) are valid
for $m=1$ and $m=n$, respectively.


\subsection{Making the domain of integration explicit}
\label{secdomainexplicit}


\subsubsection{Rewriting the indicator function}
\label{secindicator}

From the elementary observation that
\beq
R_{m-1}=S_m\cos\beta_{m}\,, \qquad R_{m}=S_m\cos\gamma_m
\label{relRmSm}
\eeq 
we deduce the relation
\begin{equation}
\frac{R_m\phantom{-}}{R_{m-1}}=\frac{\cos\gamma_m}{\cos\beta_m}\,,
\qquad m=1,\ldots,n.
\label{exprratioRR}
\end{equation}
Hence, when $R_1,\ldots,R_n$ are given, $\beta_m$ determines $\gamma_m$
and {\it vice versa}. 
We now turn to the inequality (\ref{condition2}) and divide it by
$R_m\sin\xi_m\sin\xi_{m+1}$. 
The ratios of the midpoint distances 
that then appear may be eliminated: we express
them in terms of the angular variables by means of relation
(\ref{exprratioRR}).
After some trigonometry 
$\xi_{m}$ and $\xi_{m+1}$ cancel out of the inequality, which takes the form 
$\tan\gamma_m+\tan\beta_{m+1} > 0$ for $m=1,2,\ldots,n$.
Since $-\frac{\pi}{2}<\gamma_m,\beta_{m+1}<\frac{\pi}{2}$, 
it follows that
the much simpler and geometrically necessary positivity condition
\begin{equation}
\gamma_m+\beta_{m+1} > 0,  \qquad    m=1,2,\ldots,n,
\label{cond2gammabeta}
\end{equation}
is {\it equivalent to condition\,} (\ref{condition2}).  
Therefore (\ref{cond2gammabeta}) 
defines the domain of integration in phase space.


\subsubsection{Transforming from $\xi_m$ to $\beta_m$ and $\gamma_m$}
\label{secintroducegamma}

Next, considering the outer integration
variables $R_1,\ldots,R_n$ in (\ref{exprRxipn}) as simple parameters, 
we will transform the integrals on the angles $\xi_m$ 
in two successive steps.
First, we consider $\gamma_m$ as a function of $\beta_m$
defined by (\ref{exprratioRR}), so that in view of 
(\ref{relxietabetagamma}a) we have
$\xi_m=\beta_m+\arccos[ (R_m/R_{m-1})\cos\beta_m ]$.
We transform from $\xi_m$ to the new integration variable $\beta_m$,
taking due account of
the Jacobian $j_m=\dd\xi_m/\dd\beta_m$\,, which is equal to
\beq
j_m = \frac{\sin(\beta_m+\gamma_m)}{\cos\beta_m\sin\gamma_m}.
\label{defjacxibeta}
\eeq
We order the $\beta_m$ integrations such that, going from the 
outermost one inward, we encounter successively 
those on $\beta_1,\ldots,\beta_n$.
We integrate $\beta_1$ on its full interval
$(-\frac{\pi}{2},\frac{\pi}{2})$. 
For given $\beta_1$ equation (\ref{exprratioRR}) 
fixes $\gamma_1$. Inside the $\beta_1$ integral, 
in order to satisfy condition (\ref{cond2gammabeta}) for $m=1$,
we integrate $\beta_2$ from $-\gamma_1$ to $\frac{\pi}{2}$. 
For given $\beta_2$ equation
(\ref{exprratioRR})  fixes $\gamma_2$, and so on.
In this way the conditions (\ref{cond2gammabeta}) are all
satisfied, except the one at the end which
requires $\gamma_n+\beta_1>0$ and which we will therefore
incorporate below by means of a Heaviside theta function.
After all this equation (\ref{exprRxipn}) becomes
\bea
{p}_n &=& \frac{2\pi}{n}
\int_0^\infty\!\!R_1\dd R_1\ldots R_n\dd R_n 
\int_{-\pi/2}^{\pi/2}\!\!\dd\beta_1
\int_{-\gamma_{_1}}^{\pi/2}\!\!\dd\beta_2\ldots
\int_{-\gamma_{_{n-1}}}^{\pi/2}\!\!\dd\beta_n  \nonumber\\
&& \times\,\,
\theta(\gamma_n+\beta_1)\,j_1 j_2 \ldots j_n\,
\delta\big(\sum_{m=1}^n(\beta_m+\gamma_m)-2\pi\big)\,\,
\ee^{-{\cal A}},
\label{exprRbetapn}
\eea
in which $\gamma_m$, wherever it occurs, is a function of 
$\beta_m$ given by (\ref{exprratioRR}), and where
the indicator $\chi$ is no longer needed. 
The passage from (\ref{exprRxipn}) to (\ref{exprRbetapn}) has
broken the symmetry between the $\beta_m$ and the $\gamma_m$ 
and introduced an orientation in the cell perimeter.
We have moreover broken the cyclic invariance in the index $m$ 
by imposing a fixed order of integration of the angular variables.
These symmetry breakings first appeared in an initial approach
\cite{HJHunpublished} where we constructed the
perimeter segments by a Markov process with transition probabilities between
successive vertices. We will not elaborate on this here.

We wish to introduce in (\ref{exprRbetapn})
the $\gamma_1,\gamma_2,$ $\ldots,\gamma_n$ as 
additional independent variables of integration.
One easily verifies that
\beq
\int_{-\beta_m}^{\pi/2}\!\dd\gamma_m\,\, 
\frac{\sin\gamma_m}{\cos\beta_m}R_{m-1}\, 
\delta \Big(R_m-\frac{\cos\gamma_m}{\cos\beta_m}R_{m-1}\Big)  \,=\,1.
\label{verifydelta}
\eeq
We now take the product on $m$ of the LHS of this expression 
and insert it in the integrand of (\ref{exprRbetapn}).
Still defining \cite{footnote2} 
\beq
T_m=\frac{\sin(\beta_m+\gamma_m)}{\cos^2\beta_m}
\label{defTm}
\eeq
we find that (\ref{exprRbetapn}) becomes 
\bea
{p}_n &=& \frac{2\pi}{n}
\int_0^\infty\!\!R_1\dd R_1\ldots R_n\dd R_n 
\int\dd\beta\dd\gamma\,\,
\nonumber\\
{}&& \times\,
\Big[ \prod_{m=1}^n T_m R_{m-1}\,
\delta\big(R_m-\frac{\cos\gamma_m}{\cos\beta_m}R_{m-1}\big) \Big]\,\,
\ee^{-{\cal A}}
\label{exprbetagammapn}
\eea
with our earlier convention $R_0 \equiv R_n$ 
and where we employ the shorthand notation
\bea
\int\dd\beta\dd\gamma &=&
\int_{-\pi/2}^{\pi/2}\!\!\dd\beta_1
\int_{-\beta_{_1}}^{\pi/2}\!\!\dd\gamma_1 
\int_{-\gamma_{_1}}^{\pi/2}\!\!\dd\beta_2
\int_{-\beta_{_2}}^{\pi/2}\!\!\dd\gamma_2 
\ldots\int_{-\gamma_{_{n-1}}}^{\pi/2}\!\!\dd\beta_n 
\int_{-\beta_{_n}}^{\pi/2}\!\!\dd\gamma_n \nonumber\\
&& \times\,     
\theta(\gamma_n+\beta_1)\,
\delta\big(\sum_{m=1}^n(\beta_m+\gamma_m)-2\pi\big). 
\label{defintbetagamma}
\eea
The operator $\int\!\dd\beta\dd\gamma$ represents the integral on all
possible shapes of the Voronoi cell without any reference to its radial
scale. The fact that the domain of integration appears here
fully explicitly constitutes the main achievement of this subsection.
\vspace{3mm}

An expression for ${\cal A}$ in
terms of the variables of integration $R_m$, $\beta_m$, and $\gamma_m$
may be obtained directly from its geometrical
definition as the area of the union of $n$ disks.
After the necessary rewriting one finds 
\beq
{\cal A} = \tfrac{1}{4} \sum_{m=1}^n R_m^2\,
\big[ a(\gamma_m) + a(\beta_{m+1}) \big],
\label{exprAa}
\eeq
where $a$ is defined by
\beq
a(\omega) = \omega + \tan\omega + \omega \tan^2\omega, 
\label{defalpha}
\eeq
which is equivalent, and close in form,
to the expression used by Drouffe and Itzykson \cite{DI84}
in their simulations.


\subsection{Radial integrations}
\label{secradint}


\subsubsection{Introducing the mean midpoint distance $R_{\rm av}$}
\label{secmeanmidpoint}

The average value of the $n$ midpoint distances  
$R_1,R_2,$ $\ldots, R_n$ will be denoted by 
\beq
R_{\rm av} = n^{-1}\sum_{m=1}^nR_m\,.
\label{defR}
\eeq
The circle of radius $R_{\rm av}$ will act as a 
circle of reference and we 
proceed on the hypothesis that the area
difference ${\cal A}-\pi R_{\rm av}^2$ will be small in a sense still to be
determined \cite{footnote3}. 
Let us now normalize the $R_m$ according to
\beq
R_m = R_{\rm av}\rho_m\,,
\label{defrhom}
\eeq
so that because of (\ref{defR}) the `reduced midpoint distances' 
$\rho_m$ satisfy
\beq
n^{-1}\sum_{m=1}^n\rho_m = 1.
\label{sumrulerho}
\eeq
Instead of (\ref{defrhom}) other choices of normalization would have been
possible. Therefore we will henceforth refer to (\ref{sumrulerho}) and
equivalent relations as the `gauge' condition; this in
order to distinguish it from other sum rules that play a role. 

We insert a factor
$\delta(R_{\rm av}-n^{-1}\sum_{m=1}^nR_m)$
in the integrand of the RHS of (\ref{exprbetagammapn})
and apply an extra integration on $R_{\rm av}$ to it.
The radial integrations then take the form
\bea
\int_0^\infty\!\!\dd R_1\ldots\dd R_n &=&
\int_0^\infty\!\!\dd R_{\rm av} \int_0^\infty\!\! \dd R_1\ldots\dd R_n\,\,
\delta(R_{\rm av}-n^{-1}\sum_{m=1}^n R_m) \nonumber\\
&=& \int_0^\infty\!\! \dd R_{\rm av}\, R_{\rm av}^{n-1} 
\int_0^\infty\!\! \dd \rho_1 \ldots\dd \rho_n\,\,
\delta\big( 1-n^{-1}\sum_{m=1}^n\rho_m \big)\nonumber\\
&&
\label{introRrhom}
\eea
and as a result (\ref{exprbetagammapn}) becomes
\bea
{p}_n &=& \frac{2\pi}{n} \int_0^\infty\!\! \dd R_{\rm av}\, R_{\rm av}^{2n-1}
\int_0^\infty\!\! \dd\rho_1\ldots\dd\rho_n\,\,
\delta\big( 1-n^{-1}\sum_{m=1}^n\rho_m \big) \nonumber\\
&& \times\,
\int \dd\beta\dd\gamma\,\,
\Big[ \prod_{m=1}^n
\delta\big(\rho_m-\frac{\cos\gamma_m}{\cos\beta_m}\rho_{m-1}\big)
\Big]\,
\Big[ \prod_{m=1}^n \rho_m^2 T_m^{\phantom{-1}} \Big]
\ee^{-{\cal A}},\phantom{xx}
\label{exprrhobetagammapn}
\eea
where by convention $\rho_0=\rho_n$ and
in which the angular integrations are defined by (\ref{defintbetagamma}).


\subsubsection {Integrating on the $R_m$ at 
fixed $R_{\rm av}$}
\label{secintonRm}

The $n-1$
integrations on $\rho_1,\ldots,\rho_{n-1}$ may be carried out successively
and, due to the delta functions in the integrand, cause 
$\rho_m$, wherever it occurs, to be replaced by
\beq
\rho_m=\frac{\cos\gamma_m\cos\gamma_{m-1}\ldots\cos\gamma_1}
{\cos\beta_m\cos\beta_{m-1}\ldots\cos\beta_1}\, \rho_n\,,
\qquad m=1,\ldots,n-1.
\label{condcosbcosg}
\eeq
There remains one delta function inside the product on $m$
in (\ref{exprrhobetagammapn}). We rewrite it as
\beq
\delta \Big (\rho_n 
-\frac{\cos\gamma_1\ldots\cos\gamma_n}{\cos\beta_1\ldots\cos\beta_n}
\rho_n \Big) = 
\frac{\delta(G)}{2\pi\rho_n},
\label{transfdelta}
\eeq
where
\beq
G = \frac{1}{2\pi} 
\sum_{m=1}^n\big( \log\cos\gamma_m - \log\cos\beta_m \big).
\label{deff}
\eeq
The integration on $\rho_n$ in (\ref{exprrhobetagammapn}), finally,
is easily performed due to the delta function
constraint in the first line of that equation and
leads to the cancellation of the factor $\rho_n^{-1}$ in
(\ref{transfdelta}). 
The result is that (\ref{exprrhobetagammapn}) becomes
\bea
{p}_n &=& \frac{1}{n}\int \dd\beta\dd\gamma\,\, 
\delta(G)\,
\Big[ \prod_{m=1}^n \rho_m^2 T_m \Big]\,\, 
\int_0^\infty \!\dd R_{\rm av}\, R_{\rm av}^{2n-1}\,\,
\ee^{-{\cal A}},
\label{exprbgpn}
\eea
in which now the reduced midpoint distances
$\rho_1,\ldots,\rho_n$ should be viewed as 
functions of the integration variables
determined by the $n-1$ equations (\ref{condcosbcosg}) together with 
condition (\ref{sumrulerho}).

It appears from (\ref{exprbgpn}) that
the $2n$ angles of integration
satisfy a nontrivial supplementary constraint $G=0$.
This constraint is easily interpreted: 
for an arbitrary set of angles 
$(\beta,\gamma)\equiv\{\beta_m,\gamma_m\,|\,m=1,\ldots,n\}$ 
the perimeter will spiral instead of
close onto itself after a full turn of $2\pi$;
the factor $\delta(G)$ enforces its closure.
We will refer to it as the `no-spiral' constraint.
We see that equation (\ref{transfdelta}) imposes that (\ref{condcosbcosg})
hold for $m=n$.
Hence requiring the validity of (\ref{condcosbcosg}) for 
all $m=1,\ldots,n$ is
another way of incorporating the no-spiral constraint.

In what follows we will extensively use an auxiliary variable 
$\tau_m$ representing the excess length of the reduced
midpoint distances $\rho_m$ over
their mean value, 
\beq
\rho_m=1+\tau_m\,, \qquad m=1,\ldots,n.
\label{deftau}
\eeq
From (\ref{sumrulerho}) one deduces that they satisfy 
\beq
\sum_{m=1}^n\tau_m=0,
\label{sumruletau}
\eeq
which is an alternative expression for the gauge condition.


\subsubsection{Integrating on the mean midpoint distance $R_{\rm av}$}
\label{secintonR}

Equations (\ref{defrhom}), (\ref{sumrulerho}), and (\ref{condcosbcosg}) 
allow us to express $R_m$ as  $R_{\rm av}$ times an $m$ dependent function 
of the angles which, if desired, may be made explicit.
It is then clear from (\ref{exprAa}) that  
${\cal A}$ is $R_{\rm av}^2$
times a function only of the angles. 
Bearing in mind that we expect ${\cal A}$ to be close to the area $\pi
R_{\rm av}^2$ enclosed by the circle of reference, and
anticipating later developments, we write 
\beq
{\cal A}=\pi R_{\rm av}^2\,(1+n^{-1}V),
\label{defV}
\eeq
which defines $V$.
Integrating on $R_{\rm av}$ then converts (\ref{exprbgpn}) into 
\bea
{p}_n &=& \frac{(n-1)!}{2n} \int\!\dd\beta\dd\gamma\,\,
\delta(G)\, 
\Big[ \prod_{m=1}^n \rho_m^2 T_m^{\phantom{2}} \Big]\,\, 
[\pi(1+n^{-1}V)]^{-n}.
\label{exprVpn}
\eea
The integrations that remain in the RHS of (\ref{exprVpn})  
bear exclusively on the angular variables, {\it i.e.} on the shape
of a cell with average midpoint distance scaled to unity.

For completeness and for use 
in section \ref{secexprmathV}
we now determine $V$ explicitly.
Substituting (\ref{deftau}) in
(\ref{defrhom}) and subsequently (\ref{defrhom}) 
and (\ref{defalpha}) in (\ref{exprAa}), 
comparing to (\ref{defV}), and still using 
the geometrical sum rule (\ref{sumrulebetagamma}), we obtain 
\bea
V(\beta,\gamma) &=& 
\frac{n}{4\pi}\summ (1+\tau_m)^2 \big[ \tan\gamma_m-\gamma_m 
+\tan\beta_{m+1}-\beta_{m+1} \nonumber\\
&& \phantom{xxxxxxxxxxxxxx} + \gamma_{m}\tan^2\gamma_{m}
   + \beta_{m+1}\tan^2\beta_{m+1} \big] \nonumber\\
&& +\, \frac{n}{2\pi}
   \summ (2\tau_m+\tau_m^2)(\gamma_m + \beta_{m+1}).
\label{exprVbetagammatau}
\eea
We note that (\ref{exprVbetagammatau})
is antisymmetric under a change of sign of all the angles. 


\subsubsection{The probability distribution of $R_{\rm av}$}
\label{secdistributionRav}

In this subsection we sidestep and make
an observation that will be exploited only in section
\ref{secLewis}. 
Although the outcome (\ref{exprVpn}) of the integral
on $R_{\rm av}$ is known, something more may be learned by
applying a steepest descent analysis to it.
Assuming, as is suggested by the notation $n^{-1}V$ in (\ref{defV}) and as
will be confirmed in section \ref{secfactor1pV}, that
$V(\beta,\gamma)$ remains of order $n^0$
when $n\to\infty$, we then find 
that the integral on $R_{\rm av}$ draws its main contribution from a Gaussian
centered at the saddle point $R_{{\rm av},c}$ and having a
variance $\sigma_{{\rm av},c}^2$ given by
\bea
R_{{\rm av},c}(\beta,\gamma) &=& 
\Big[ \frac{2n-1}{2\pi(1+n^{-1}V)} \Big]^{\half}
=  \Big( \frac{n}{\pi} \Big)^\half \big( 1+{\cal O}(n^{-1}) \big),
\label{exprRav} \\[2mm]
\sigma_{{\rm av},c}^2 (\beta,\gamma)&=& [4\pi(1+n^{-1}V)]^{-1} 
= \frac{1}{4\pi} \big( 1+{\cal O}(n^{-1}) \big).
\label{exprvarRav}
\eea
The results (\ref{exprRav}) and (\ref{exprvarRav}) 
are to leading order in $n$ independent of $(\beta,\gamma)$.
The ${\cal O}(n^{-1})$ corrections in these equations
do depend on $(\beta,\gamma)$ but will play no role in this work.
Hence, provided we confirm the scaling of $V$, we have found here that
the $n$-\-sided Voronoi cell has a
characteristic radial scale of order $n^\half$ \cite{footnoteimplications}. 
Furthermore, if we define
\bea
R_c = \pi^{-\half}n^\half, \qquad 
R_{\rm av}=R_c( 1+n^{-\half}r_{\rm av}),
\label{defrav}
\eea
then it follows from the above that in the limit $n\to\infty$ the variable
$r_{\rm av}$ has the probability distribution
\beq
p_{\rm av}(r_{\rm av})=2^\half\pi^{-\half}
\exp\big(\!-2r_{\rm av}^2\big) + {\cal O}(n^{-\half}).
\label{defprav}
\eeq
With $\deltaR_{\rm av}=n^{-\half}R_cr_{\rm av}$ and with
the midpoint density $4\lambda$ restored this becomes equation
(\ref{defpRav}) of the introduction. 


\subsection{Passing to the angles of integration $\xi_m$ and $\eta_m$}

So far the $\beta_m$ and $\gamma_m$ [the set (\ref{defsetbetagamma})] 
have been the variables most convenient to work
with. However, in order to prepare for the large $n$ expansion, it is now
necessary that we pass to the set of variables
(\ref{defsetxieta})  consisting of the
$\xi_m$ and $\eta_m$ as well as of $\beta_1$.
Employing the notation $(\xi,\eta)=\{\xi_m,\eta_m\,|\,m=1,\ldots,n\}$,
$\xi=\{\xi_m\,|\,m=1,\ldots,n\}$, and $\eta=\{\eta_m\,|\,m=1,\ldots,n\}$,
we now proceed as follows.

By writing $G(\xi,\eta;\beta_1)$ we indicate explicitly that we
want to consider $G$ as a function of $\xi$, $\eta$, and $\beta_1$.
The prime on $G'$ stands for
differentiation with respect to $\beta_1$.
Let furthermore $\beta_*(\xi,\eta)$ denote 
the solution of the no-spiral constraint
\beq
G(\xi,\eta;\beta_*)=0.
\label{Gbetastar0}
\eeq 
In (\ref{exprVpn}) one easily converts
the integration on $(\beta,\gamma)$, 
defined in (\ref{defintbetagamma}), 
to one on the new variables $(\xi,\eta)$
by shifting the integration intervals appropriately
(except the outermost one, for $\beta_1$, which remains unchanged)
and employing the fact that 
$\gamma_n+\beta_1=2\pi-\sum_{m=1}^{n-1}\eta_m$.
With the aid of the identity
\beq
\theta\big(2\pi-\sum_{m=1}^{n-1}\eta_m\big) 
= \int_0^{2\pi}\!\!\dd\eta_n\,
\delta \big( \sum_{m=1}^{n}\eta_m - 2\pi \big)
\label{relthetadelta}
\eeq
and using the occasion to rearrange some factors in the integrand
we recast (\ref{exprVpn}) in the form
\bea
{p}_n &=& 
\frac{(n-1)!}{2n}\int_{-\pi/2}^{\pi/2}\!\!\dd\beta_1
\int\!\!\dd\xi\dd\eta\,\,
\delta\big(\sum_{m=1}^n\xi_m-2\pi\big)\,
\delta\big(\sum_{m=1}^n\eta_m-2\pi\big) \nonumber\\[2mm]
&&\times\,
\frac{\delta ( \beta_1-\beta_* )}{|G'(\xi,\eta;\beta_*)|}\,
\Big[\prod_{m=1}^n \rho_m^2 T_m^{\phantom{2}} \Big]
[\pi(1+n^{-1}V)]^{-n} 
\label{exprhatPsigma}
\eea
where
\bea
\int\dd\xi\dd\eta &=&
\int_0^{\pi/2+\beta_{_1}}\!\!\dd\xi_1 
\int_0^{\pi/2+\gamma_{_1}}\!\!\dd\eta_1
\int_0^{\pi/2+\beta_{_2}}\!\!\dd\xi_2 \ldots \nonumber\\[1mm]
&& \phantom{xxxxxxx}
\ldots\int_0^{\pi/2+\gamma_{_{n-1}}}\!\!\dd\eta_{n-1} 
\int_0^{\pi/2+\beta_{_n}}\!\!\dd\xi_n\,\,     
\int_0^{2\pi}\dd\eta_n\,.\phantom{xx}
\label{defintxieta}
\eea
The notation is hybrid; the variables
$\gamma_1,\beta_2,\gamma_3,\ldots,\beta_n$ 
in the upper integration limits in (\ref{defintxieta})
should be viewed as functions of $\xi$, $\eta$, and $\beta_1$ 
given by (\ref{inversebgxy}). 
Each limit of integration
depends only on the variables of more outward integrations, as of
course it should be.

We note that it is not possible at this stage to carry out the integral
on $\beta_1$, since $\beta_*$ depends on $\xi$ and $\eta$ and the
integration limits of the integrals $\int\dd\xi\dd\eta$ in turn
depend on $\beta_1$.


\subsection{Free and interacting problem}
\label{secLTorders}


\subsubsection{Defining a free problem and an interaction}
\label{secLT}

In order to relax the constraints 
(\ref{sumrulexi}) and (\ref{sumruleeta})
enforced by the first two delta
functions in (\ref{exprhatPsigma}), 
we will exploit the integral representation
$\delta(X)=\int_{c-{\rm i}\infty}^{c+{\rm i}\infty}
\!\dd s\,\exp(-sX)$, valid for any constant $c$.
Upon introducing a parameter $\kappa$ which at this point is
arbitrary and scaling conveniently with $n$ we can write
\bea
&&\delta\big(\tsum_m\xi_m-2\pi\big)
\delta\big(\tsum_m\eta_m-2\pi\big)= 
\label{Fourierrepr}
\nonumber\\[2mm]
&&\phantom{xx}
= \delta\big(\kappa\tsum_m\xi_m+(1-\kappa)\tsum_m\eta_m-2\pi\big)\,
\delta\big(\tsum_m\xi_m-\tsum_m\eta_m\big) \\[2mm]
&&\phantom{xx}
= \Big(\frac{n}{2\pi{\rm i}}\Big)^{\!2}\!
\int_{c-{\rm i}\infty}^{c+{\rm i}\infty}\!\!\dd s
\int_{-{\rm i}\infty}^{{\rm i}\infty}\!\!\dd t\,\,
\ee^{2\pi ns-ns[\kappa\sum_m\xi_m
-(1-\kappa)\sum_m\eta_m]-nt(\sum_m\xi_m-\sum_m\eta_m)},\nonumber 
\eea
where we choose $c>0$. Substituting (\ref{Fourierrepr})
in (\ref{exprhatPsigma}) we find \cite{footnote4}
\bea
p_n&=&
\Big(\frac{n}{2\pi{\rm i}}\Big)^{\!2}\,
\int_{c-{\rm i}\infty}^{c+{\rm i}\infty}\!\dd s 
\int_{-{\rm i}\infty}^{{\rm i}\infty}\!\dd t
\int_{-\pi/2}^{\pi/2}\!\dd\beta_1\,\,
\ee^{2\pi ns} 
\int\!\dd\xi\dd\eta\,\delta(\beta_1-\beta_*)\,
\ee^{-{\mathbb H}(s,t)}\nonumber\\[2mm]
&\equiv& {\rm Tr}\,\ee^{-{\mathbb H}(s,t)},
\label{FTinv}
\eea
in which the last line defines the \,Tr\, operator, and
the `Hamiltonian' ${\mathbb H}$ is given as the sum of a `free' Hamiltonian
${\mathbb H}_{\,0}$ and an `interaction' ${\mathbb V}$,
\beq
{\mathbb H}(s,t) = {\mathbb H}_{\,0}(s,t)\,+\,{\mathbb V},
\label{defsplitmathbbH}
\eeq
where
\bea
\ee^{-{\mathbb H}_{\,0}(s,t)} &=& \pi^{-n}\Big[
\prod_{m=1}^n 
\xi_m\, \ee^{-ns[\kappa\xi_m+(1-\kappa)\eta_m]-nt(\xi_m-\eta_m)} \Big], 
\label{defmathbbH0} \\
\ee^{-{\mathbb V}} &=& |G'(\xi,\eta;\beta_*)|^{-1}
\Big[ \prod_{m=1}^n \rho_m^2 T_m^{\phantom{2}} \xi_m^{-1} \Big]
(1+n^{-1}V)^{-n}.
\label{defmathbbV}
\eea
The factor $\xi_m$ inserted in (\ref{defmathbbH0})
and compensated in (\ref{defmathbbV}) serves to
ensure that $T_m^{\phantom{2}}\xi_m^{-1}$ 
has a finite limit as $\xi_m\to 0$.
The Hamiltonian ${\mathbb H}_{\,0}$ defined by (\ref{defmathbbH0})
is obviously is sum of $2n$ independent single-variable terms. 
However, 
the reason why the splitting (\ref{defsplitmathbbH}) should be exactly as
defined by (\ref{defmathbbH0}) and (\ref{defmathbbV}), 
and not something else, will have to be borne out by the calculation.

With equations (\ref{FTinv})-(\ref{defmathbbV}) 
we have achieved the purpose of this section.
They have nothing of the simple elegance of other
model Hamiltonians in statistical mechanics and are at first sight 
hardly suggestive of any particular analytical approach.
However, the work of this section has served to 
prepare the ground for what follows.

So far our calculation is valid for all finite $n=3,4,\ldots$\,. 
The initial problem (\ref{exprR2pn})
is exactly equivalent to problem (\ref{FTinv}).
However, we cannot continue beyond the present point: 
it is impossible to carry out the integral on 
$\beta_1$ in (\ref{FTinv}) since $\beta_*$ depends on 
the $\xi_m$ and $\eta_m$
whose upper integration limits in turn
depend on $\beta_1$. This entanglement
will be lifted in section
\ref{secscaling} as soon as we start considering the limit $n\to\infty$.


\subsubsection{Splitting the calculation}
\label{secorders}

Ensuant on equation (\ref{defsplitmathbbH}) we have the formal factorization
\beq
{\rm Tr\,} \ee^{-{\mathbb H}}=
\big({\rm Tr\,} \ee^{-{\mathbb H}_{\,0}}\big)\,
\langle\ee^{-\mathbb{V}}\rao\,,
\label{deforders}
\eeq
whence follows the splitting
\beq
\log p_n\,=\,\log p_n^{(0)} 
\,+\,\log\langle\ee^{-\mathbb{V}}\rao\,,
\label{splitlogpn}
\eeq
where we employ the alternative notation 
$p_n^{(0)}={\rm Tr}\,\ee^{-{\mathbb H}_{\,0}}$. 
We will refer to the calculation of $p_n^{(0)}$ as the 
`free' problem and to that of $\langle\ee^{-\mathbb{V}}\rao$
as the `interacting problem.'
Although technically not strictly necessary,
it will greatly enhance the clarity of presentation
to treat these problems in two successive steps.

In the course of our work we will also be interested in 
correlation functions of the angles.
We write $\la F\rao$ for the `noninteracting' average of any function $F$ 
of the angular coordinates, {\it i.e.,} 
\beq
\langle F\rao \,=\, \frac {{\rm Tr}\,F\,\ee^{-{\mathbb H}_{\,0}}} 
{{\rm Tr}\,\ee^{-{\mathbb H}_{\,0}}}. 
\label{defFaverage}
\eeq
A full average $\la F\ra$ may 
be reexpressed in terms of noninteracting
averages according to
\beq
\la F\ra = \frac{{\rm Tr\,}F\,\ee^{-{\mathbb H}}}
{{\rm Tr\,}\ee^{-\mathbb H}}
= \frac{\la F\,\ee^{-{\mathbb V}}\rao}{\la\ee^{-{\mathbb V}}\rao}.
\label{averagesF}
\eeq

In section \ref{seczerothorder} we will 
solve the noninteracting problem in the limit $n\to\infty$.
The study of the interaction $\mathbb V$ will be taken up in
section \ref{secexprmathV}.
It will require considerable work to show
that the second term in (\ref{splitlogpn}),
$\log \la\ee^{-\mathbb{V}}\rao$, 
has a well-controlled expansion in powers of $n^{-\half}$.
Correlation functions will be considered in sections
\ref{secV} and \ref{secdeviations}.


\section{Scaling}
\label{secscaling}

\subsection{Scaling of $\xi_m$ and $\eta_m$ for large $n$}
\label{secscalingxy}

In the preceding section, where $n$ was an arbitrary integer, 
we have established the starting point for an expansion for
asymptotically large $n$.
We will now make the first steps in this expansion.
A major problem is to know in advance how the various quantities
will scale with $n$. 
For $n\to\infty$
we expect all angles to become small and we begin by introducing
\beq
\xi_m=n^{-1}x_m, \qquad \eta_m=n^{-1}y_m, \qquad m=1,\ldots,n.
\label{scxieta}
\eeq
This scaling of $\xi_m$ and $\eta_m$ is certainly expected, since they
constitute two sets of
$n$ positive variables adding up to $2\pi$. 


\subsection{Scaling the integral for $p_n$}
\label{secconsequences}

The integral $\int\!\dd\xi\dd\eta$ included in the trace
in (\ref{FTinv})
is defined in (\ref{defintxieta}). It there appears that 
the upper integration limits of the $\xi_m$ and $\eta_m$,
given in terms of the $\beta_\ell$ and $\gamma_\ell$, 
are all positive.
Hence after passing to the 
scaled variables $x_m$ and $y_m$ we may replace these 
upper integration limits by $\infty$. Since the integrals converge
exponentially,
the error should be expected to
vanish exponentially with $n$. 
It then becomes  possible to commute the
integration on $\beta_1$ through those on
the $x_m$ and $y_m$ and carry it out with the aid of the factor
$\delta(\beta_1-\beta_*)$. Henceforth the angle $\beta_1$, wherever it still
occurs (such as in expressions for ${\mathbb V}$ to be given later), 
should be viewed as the function $\beta_*(\xi,\eta)$ given by
the solution of (\ref{Gbetastar0}).

As a result the integral (\ref{FTinv}) takes the simplified form 
\bea
p_n&\simeq&
\Big(\frac{n}{2\pi{\rm i}}\Big)^{\!2}\,\frac{(n-1)!}{2n}\pi^{-n}n^{-3n}
\int_{c-{\rm i}\infty}^{c+{\rm i}\infty}\!\dd s 
\int_{-{\rm i}\infty}^{{\rm i}\infty}\!\dd t\,\,
\ee^{2\pi ns} \nonumber\\
&& \times
\int_0^\infty\Big[\prod_{m=1}^n\dd x_m\dd y_m\Big]
\Big[\prod_{m=1}^n x_m\ee^{-s\kappa x_m-s(1-\kappa)y_m - t(x_m-y_m)}\Big]
\nonumber\\[2mm]
&& \times\,\exp\big (-{\mathbb V}\,(x,y) \big),
\label{LTinvscaling}
\eea
where we indicated explicitly that ${\mathbb V}$, given by
(\ref{defmathbbV}), depends on
the $x_m$ and $y_m$.
We use the $\simeq$ sign to indicate that this is an asymptotic
equality.

Provided that ${\mathbb V}$ be sufficiently well-behaved, the integrals
on the $x_m$ and $y_m$ will converge at the condition that
\beq
{\rm Re\,}(\kappa s + t)>0, \qquad {\rm Re\,}((1-\kappa)s - t)>0.
\label{condkappa}
\eeq
Since $\,{\rm Re\,} t =0\,$ and $\,{\rm Re\,} s =c>0$,\, 
this condition is
satisfied if we choose $\kappa$ in the interval $0<\kappa<1$.
It is now of great advantage to transform to variables 
of integration $u$ and $v$ defined by
\beq
u=\kappa s + t, \qquad v=(1-\kappa)s-t.
\label{trfstuv}
\eeq
In terms of these equation (\ref{LTinvscaling}) becomes
\bea
p_n&\simeq&
\Big(\frac{n}{2\pi{\rm i}}\Big)^{\!2}\,\frac{(n-1)!}{2n}\pi^{-n}n^{-3n}
\int \!\dd u\, \ee^{2\pi nu} 
\int \!\dd v\, \ee^{2\pi nv} \nonumber\\[2mm]
&& \times
\int_0^\infty\Big[\prod_{m=1}^n\dd x_m\,x_m\,\ee^{-ux_m}\Big]
\int_0^\infty\Big[\prod_{m=1}^n\dd y_m\,\ee^{-vy_m}\Big]
\nonumber\\[2mm]
&& \times\,\exp\big(\! -{\mathbb V}\,(x,y) \big),
\label{LTinvscalinguv}
\eea
where $u$ and $v$ run independently from $\,-{\rm i}\infty\,$ 
to $\,{\rm i}\infty\,$ while passing to the right of the origin.
The difficulty of evaluating (\ref{LTinvscalinguv}) is now hidden in the
expression for ${\mathbb V}$.

The final remarks of this section concern practical procedure.
Since (\ref{LTinvscalinguv}) differs from (\ref{FTinv}) 
only by terms that are 
exponentially small in $n$ and since we are preparing for a
perturbation series in inverse powers of $n$,
the scaled integral
(\ref{LTinvscalinguv}) together with expression (\ref{defmathbbV})
for ${\mathbb V}$ is as good a starting point as 
the original integral (\ref{FTinv}) 
together with (\ref{defsplitmathbbH})-(\ref{defmathbbV}).
In particular, (i) setting ${\mathbb V}=0$ in in (\ref{LTinvscalinguv})
yields $p_n^{(0)}$; and (ii)
the average $\la F\rao$ defined in
(\ref{defFaverage}) is, up to exponentially small differences,
equal to expression 
(\ref{LTinvscalinguv}) with $F$ inserted and ${\mathbb V}$ set to zero, 
and divided by $p_n^{(0)}$. 


\section{Free problem}
\label{seczerothorder}


\subsection{Calculation of\,\, $p_n^{(0)}$}
\label{secPnzerothorder}

We will solve the noninteracting problem by calculating $p_n^{(0)}$
starting from expression (\ref{LTinvscalinguv}) in which we set
${\mathbb V}=0$.
The $2n$ integrals 
on the $x_m$ and $y_m$, as well as those on $u$ and $v$,
then factorize. After carrying out the former we find 
\bea
p_n^{(0)}&\simeq&
\frac{(n-1)!}{2n}\pi^{-n}n^{-3n}
\Big(\frac{n}{2\pi{\rm i}}\int\!\dd u\,\frac{\ee^{2\pi nu}}{u^{2n}}\Big)
\Big(\frac{n}{2\pi{\rm i}}\int\!\dd v\,\frac{\ee^{2\pi nv}}{v^n}\Big)
\nonumber\\[2mm]
&=&
\frac{(n-1)!}{2n}\pi^{-n}n^{-3n}\times
\frac{n^{2n}(2\pi)^{2n-1}}{(2n-1)!}\times 
\frac{n^{n}(2\pi)^{n-1}}{(n-1)!} \nonumber\\[2mm]
&=& \frac{(8\pi^2)^n}{4\pi^2(2n)!},
\label{resP0}
\eea
where in the first step we did the integrations by closing
their contours around the poles in the origin.
Additive corrections to (\ref{resP0}) 
should be expected to vanish exponentially with $n$.

It is instructive and useful for later to analyze the integrals also by
the steepest descent method. In the case of the $u$ integral the
integrand may be written as $\exp(-2n\log u + 2\pi nu)$,
which has a saddle point for $u=u_c\equiv \pi^{-1}$.
Upon taking its second derivative, it appears that contributions come
from a neighborhood of extension $\sim n^{-\frac{1}{2}}$ 
around the saddle point. 
A similar analysis holds for the $v$ integral, for which the saddle
point is at $v=v_c\equiv (2\pi)^{-1}$.
The steepest descent analysis allows us to conclude that any
function $h(u,v)$ independent of $n$ and inserted in the integrand
contributes an extra factor $h(u_c,v_c)$ to the value of the integral.
This rule will be applied in section \ref{seccalcpn}.


\subsection{Averages and correlations}
\label{secspecificav}

The averages of the scaled variables
$x_m$ and $y_m$ are necessarily in any order equal to
\beq
\la x_m\ra \equiv\bar{x} = 2\pi, \qquad 
\la y_m\ra \equiv\bar{y} = 2\pi.            
\label{resavxy}
\eeq
One obtains the noninteracting results 
for arbitrary multivariate moments of the
$x_m$ and $y_m$ by starting from (\ref{LTinvscalinguv}) with the 
appropriate insertions, setting again
${\mathbb V}=0$, and going through calculational steps 
similar to those above.
We state the results only for the second moments,
\beq
\la x_m^2 \rao = \frac{12\pi^2 n}{2n+1}, \qquad
\la y_m^2 \rao = \frac{ 8\pi^2 n}{ n+1}.
\label{rescorr}
\eeq
Due to the symmetry of the noninteracting expression 
[equation (\ref{LTinvscalinguv})
with ${\mathbb V}=0$], the two-point correlations $\la x_\ell x_m\ra$ and
$\la y_\ell y_m \ra$ must be index independent when $\ell\neq m$. 
This observation combined with the geometrical
sum rules (\ref{sumrulexi}) and (\ref{sumruleeta}) yields
\beq
\la x_\ell x_m \rao = \frac{8\pi^2 n}{2n+1}, \qquad
\la y_\ell y_m \rao = \frac{2\pi^2 n}{ n+1}, \qquad \ell\neq m.
\label{rescorrbis}
\eeq
Furthermore $\la x_\ell y_m \rao=\la x_\ell \rao \la y_m \rao$ 
for all $\ell$ and $m$.
Let us define
\beq
\delta x_m = x_m-\bar{x}, \qquad 
\delta y_m = y_m-\bar{y}.        
\label{defdeltaxdeltaym}
\eeq
From what precedes we then have,
still expanding (\ref{rescorr}) in powers of $n^{-1}$,
\bea
\la \delta x_\ell \delta x_{m}\rao &=& 2\pi^2 
\big( \delta_{\ell m}-n^{-1} \big)
\big( 1+{\cal O}(n^{-1}) \big),
\nonumber \\[2mm]
\la \delta y_\ell \delta y_{m}\rao &=& 4\pi^2
\big( \delta_{\ell m}-n^{-1}\big) 
\big( 1+{\cal O}(n^{-1}) \big),
\nonumber\\[2mm]
\la \delta x_\ell \delta y_{m}\rao &=& 0.
\label{exprcorrdeltaxdeltay}
\eea
These variances obey the geometrical sum rules
(\ref{sumrulexi}) and (\ref{sumruleeta}).
But although they include terms of order
$n^{-1}$, we will see in sections \ref{secV} and \ref{secdeviations} 
that the interaction ${\mathbb V}$ generates additional 
order $n^{-1}$ contributions. 


\subsection{Probability distribution of $\xi_m$ and $\eta_m$}
\label{seczerothorderaverages}

The probability for the angle $\xi_m$ to be between $\xi$ and $\xi+\dd\xi$
is given by $\la\delta(\xi_m-\xi)\ra\dd\xi$.
We set $\xi=n^{-1}x$ and Fourier represent according to
\beq
\delta(x_m-x)=\frac{1}{2\pi}\int_{-\infty}^{\infty}
\!\dd\mu\,\,\ee^{{\rm i}\mu(x_\ell-x)}.
\label{FTdelta}
\eeq
We insert this in (\ref{LTinvscalinguv}), commute the $\mu$
integral to the outside, and
pass through the same steps that led from (\ref{LTinvscalinguv})
to the first line of (\ref{resP0}). 
The result is similar to (\ref{resP0}) 
but with an insertion $u^2(u-{\rm i}\mu)^{-2}$, 
\bea
&&p_n^{(0)}\la\delta(x_m-x)\rao \simeq
\frac{1}{2\pi}\int_{-\infty}^{\infty}
\!\dd\mu\,\,\ee^{-{\rm i}\mu x}
\label{P0intuvdelta}\\[2mm]
&& \phantom{xx}\times\,\frac{(n-1)!}{2n}\pi^{-n}n^{-3n}\,
\Big(\frac{n}{2\pi{\rm i}}\int\!\dd u\, 
\frac{\ee^{2\pi nu}}{u^{2n}}\,\big(\frac{u}{u-{\rm i}\mu}\big)^2 \Big)
\Big(\frac{n}{2\pi{\rm i}}\int\!\dd v\, \frac{\ee^{2\pi nv}}{v^n}\Big).
\nonumber
\eea
Because of our discussion on insertions at the end of section  
\ref{secPnzerothorder},  
the second line in (\ref{P0intuvdelta}) is equal to
$p_n^{(0)}(u_c/(u_c-{\rm i}\mu))^2$.   
Upon dividing by $p_n^{(0)}$ and setting $u_c=1/\pi$ we 
obtain the probability distribution of $x$ in the limit $n\to\infty$,
\bea
\la\delta(x_m-x)\rao &=&
\frac{1}{2\pi}\int_{-\infty}^{\infty}\!\dd\mu\,\,
\frac{\ee^{-{\rm i}\mu x}}{(1-{\rm i}\pi\mu)^2} \nonumber\\[2mm]
&=& \frac{x}{\pi^2}\,\exp\big(\!-\frac{x}{\pi} \big).
\label{deltaximxi}
\eea
In terms of the original variable
$\xi=n^{-1}x$ this becomes the distribution $u(\xi)$ announced in
equation (\ref{resuxi}) of the
introduction. In a fully analogous way an insertion
$\delta(\eta_m-\eta)$ produces 
\beq
\la\delta(y_m-y)\rao = \frac{1}{2\pi}\,\exp\big(\!-\frac{y}{2\pi} \big),
\label{deltaetameta}
\eeq
from which $v(\eta)$ of equation\,(\ref{resveta}) follows.  


\section{Further scaling}
\label{secmorescaling}

\subsection{Scaling of $\beta_m$ and $\gamma_m$ for large $n$}
\label{secscalingbg}

The noninteracting calculations of section \ref{seczerothorder}
required no scaling other than the one
postulated in (\ref{scxieta}), which concerns the 
$\xi_m$ and $\eta_m$. 
We will show now how the scaling of all
further variables follows from this initial postulate {\it plus\,}
what we have learned about the independence of the $\xi_m$ and $\eta_m$. 

The first one of equations
(\ref{inversebgxy}) expresses $\beta_m-\beta_1$ as a sum
of $2m-2$ terms $\xi_\ell$ and $\eta_\ell$ which,
in the noninteracting case,
are all independent.
Since $m$ is typically of order $n$,
it follows that $\beta_m-\beta_1$ scales as
$n^{-\half}$. But since $\beta_1$ is in no way statistically
different from the other $\beta_m$, they must all scale as
$n^{-\half}$. In a similar way one deduces from the second one of
equations (\ref{inversebgxy}) that the $\gamma_m$
scale with $n^{-\half}$.
We therefore set
\beq
\beta_m=n^{-\half}b_m\,, \qquad \gamma_m=n^{-\half}c_m\,,
\qquad m=1,\ldots,n,
\label{scbetagamma}
\eeq
as well as
\beq
\beta_*(\xi,\eta)=n^{-\half}b_*(x,y).
\label{scbetastar}
\eeq
Hence the scaling factor for the angles
$\xi_m$ and $\eta_m$ is different from the one
for the angles $\beta_m$ and $\gamma_m$.
By symmetry and because of the geometrical sum rule (\ref{sumrulebetagamma})
we must have $\la\beta_m\ra=\la\gamma_m\ra=\pi/n$. 
The difference between the
scaling  of these averages
and of the typical values of $\beta_m$ and $\gamma_m$, 
which are $\sim n^{-\half}$,
can arise only because the $\beta_m$ and $\gamma_m$
may take negative values, as illustrated in figure \ref{fig2}. 

It will be useful to have at hand the scaled equivalent of
(\ref{inversebgxy}). Rendering explicit the fact that here
$\beta_1=\beta_*(\xi,\eta)$ we find
\bea
b_m &=& b_*(x,y) - n^{-\half}\sum_{\ell=1}^{m-1} (x_\ell-y_\ell),\\ 
c_m &=& -b_*(x,y) + n^{-\half}\sum_{\ell=1}^{m-1} (x_\ell-y_\ell) 
                  + n^{-\half} x_m\,, 
\label{inversebgxysc}
\eea
valid for $m=1,\ldots,n$.

The two sums $\sum_{\ell=1}^m x_\ell$\, and $\sum_{\ell=1}^m y_\ell$\,
may be viewed as the displacements of two independent random walks, 
each built up out of independent increments. 
Figure \ref{fig1} shows that these displacements describe the progression of
the polar angles of the midpoint 
vector $\bR_m$ and the vertex vector $\bS_m$, respectively. 
Hence the cell perimeter may be viewed
\cite{HJHletter05} as the
result of a construction based on two entangled random walks. 


\subsection {Cancellations of orders in $n^{-\half}$}
\label{secpreliminaries}

The nontriviality of this scaling picture comes from the fact that
various frequently occurring
combinations of the angular variables exhibit cancellations. 
For later use we collect some of 
them here, employing a hybrid notation in which
on the one hand the $b_m$ and $c_m$, and on the other hand the 
$x_m$ and $y_m$ appear. 
First of all there are the sums of odd powers of the type
\bea
\gamma_m+\beta_{m+1} &=& y_m n^{-1}, \nonumber\\[2mm]
\gamma_m^3 +\beta_{m+1}^3 &=& 
(\gamma_m+\beta_{m+1})\,
\big[ \tfrac{3}{2}(\gamma_m^2+\beta_{m+1}^2) - 
  \tfrac{1}{2}(\gamma_m+\beta_{m+1})^2 \big] \nonumber\\[2mm]
&=& \tfrac{3}{2}y_m (c_m^2+b_{m+1}^2)n^{-2} 
- \tfrac{1}{2} y_m^3 n^{-3},
\label{c3plusb3}
\eea
and analogously
\bea
\beta_m+\gamma_m &=& x_m n^{-1}, \nonumber\\[2mm]
\beta_m^3+\gamma_m^3 &=& \tfrac{3}{2}x_m (b_m^2+c_m^2) n^{-2}
- \tfrac{1}{2} x_m^3 n^{-3}.
\label{b3plusc3}
\eea
Secondly, we will encounter differences of even powers,
\bea
\beta_\ell^2-\gamma_\ell^2 &=&
(\beta_\ell+\gamma_\ell)(\beta_\ell-\gamma_\ell) \nonumber\\[2mm]
&=&x_\ell(b_\ell-c_\ell)n^{-\frac{3}{2}}, \nonumber\\[2mm]
\beta_\ell^4-\gamma_\ell^4 &=& \big[ (\beta_\ell+\gamma_\ell)^2 - 
2\beta_\ell\gamma_\ell \big] (\beta_\ell^2-\gamma_\ell^2) \nonumber\\[2mm]
&=& -2x_\ell b_\ell c_\ell (b_\ell-c_\ell) n^{-\frac{5}{2}} 
+ x_\ell^3(b_\ell-c_\ell) n^{-\frac{7}{2}}.
\label{b4minc4}
\eea
These relations are easily extended, if needed, to
sums and differences of higher powers of the angles. 


\subsection {Solution of $\,G(\xi,\eta;\beta_*)=0$}
\label{secG}

The solution $\beta_*(\xi,\eta)$ of the no-spiral constraint
$G(\xi,\eta;\beta_*)=0$ has not been needed explicitly. 
Since $\beta_*$ plays an important role,  
we will show in this subsection how it may
be obtained in the scaling limit.
The derivation also establishes its uniqueness
within the large $n$ perturbation expansion.
Finally, because of equivalence by symmetry, the study of $\beta_*=\beta_1$ 
gives us access to the statistics of all angles $\beta_m$ and $\gamma_m$.

From (\ref{deff}) we have straightforwardly
\bea
G(\xi,\eta;\beta_*) &=& \frac{1}{4\pi}\summ 
\Big( \beta_m^2-\gamma_m^2
\,+\,{\cal O}(\beta_m^4-\gamma_m^4) \Big) \nonumber\\[2mm]
&=& \frac{1}{4\pi n}\summ
\big( b_m^2-c_m^2 \big) \,+\, {\cal O}(n^{-\frac{3}{2}}),
\label{serG}
\eea
The leading order term is $\sim n^{-\half}$. Its scaling, as well as that
of the correction term, may be determined with the aid of
equations (\ref{b4minc4}) and taking into account the sum on $m$.
For convenience of notation we rewrite (\ref{inversebgxysc}) 
just for this occasion as 
$b_m=b_*+\tilde{b}_m$ and $c_m=-b_*+\tilde{c}_m$,
substitute it in the RHS of (\ref{serG}),
set $G=0$, and solve for $b_*(x,y)$ using 
the geometrical sum rule (\ref{sumrulebetagamma}). 
The result is that
\bea
b_*(x,y) &=& -\frac{1}{4\pi n^{\half}}\summ\,(\tilde{b}_m^2-\tilde{c}_m^2)
+ {\cal O}(n^{-1}) \nonumber\\[2mm]
&=& \frac{1}{4\pi n^{\frac{3}{2}} }\Big[
{2}\summ \sum_{\ell=1}^{m-1} x_m (x_\ell-y_\ell) 
+ \summ x_m^2 
\Big] + {\cal O}(n^{-1}), 
\label{solbstar}
\eea
where in the second line we returned to standard notation.
Inside the square brackets of (\ref{solbstar}) the double sum is of
order $n^{\frac{3}{2}}$ and the single sum of order $n$.
By taking the average $\la\ldots\rao$
on both sides of (\ref{solbstar}) and using  
(\ref{resavxy})-(\ref{rescorrbis})
one finds, after cancellation of the leading order terms, that 
$\la b_* \rao = \pi n^{-\half}$. 
This relation should in fact hold to all orders in $n$
and leads to $\la\beta_m\rao=\la\gamma_m\rao=\pi n^{-1}$,
as had to be the case.
The distribution of $b_*$ around its average is on a scale which is 
larger by a factor $n^\half$. Restricting ourselves now to this larger scale, 
we substitute $x_m=\bar{x}+\delta x_m$ in (\ref{solbstar}) and, 
reasoning as before, keep only the leading order. The result,
\beq
b_*(x,y) = -n^{-\frac{3}{2}} \sum_{m=1}^n m (\delta x_m - \delta y_m)
+{\cal O}(n^{-\half}),
\label{solbstarleading} 
\eeq 
shows that $b_*(x,y)$ is a weighted sum of $n$ weakly dependent
random variables and therefore, in the limit of large $n$, has 
a Gaussian distribution. Its noninteracting variance may be calculated from
(\ref{solbstarleading}) and (\ref{exprcorrdeltaxdeltay}) 
and is given by $\la b_*^2 \rao \simeq \half\pi^2$.
However, the interaction ${\mathbb V}$ yields
an additional ${\cal O}(n^0)$ contribution to this variance.
Its explicit value to order $n^0$ may be found
analytically with the aid of the interacting theory of the
next sections, but we will not pursue this calculation. 


\section{Interaction}
\label{secexprmathV}

The total Hamiltonian 
${\mathbb H}$ was split in equation (\ref{defsplitmathbbH}) 
into a noninteracting
term ${\mathbb H}_{\,0}$ and an interaction ${\mathbb V}$,
with an ensuing splitting for $\log p_n$ in (\ref{splitlogpn}).
In section \ref{seczerothorder} we dealt with the noninteracting problem
and obtained, in equation (\ref{resP0}), 
the first term of (\ref{splitlogpn}).
This result is useful only if we can also control the
second term in (\ref{splitlogpn}), which requires the evaluation of 
$\la\ee^{-{\mathbb V}}\rao$. Furthermore, if we wish to study correlations,
$\la F\,\ee^{-{\mathbb V}}\rao$ is needed.
We will now consider these quantities.


\subsection{General}
\label{secVgeneral}

The interaction ${\mathbb V}$ 
is defined by expression (\ref{defmathbbV}), which is particularly opaque.
We now assert that ${\mathbb V}$ it is in fact of order $n^0$.
We will show this by constructing an analytically tractable expression 
${\mathbb V}_1$ such that \cite{footnote5} 
\beq
{\mathbb V} = {\mathbb V}_1 + n^{-\half}{\mathbb V}_2
\label{seriesmathbbV}
\eeq
with ${\mathbb V}_1$ of order $n^0$ and 
$n^{-\half}{\mathbb V}_2$ representing a collection of higher order terms.
Expanding in a power series in $n^{-\half}{\mathbb V}_2$ we obtain from
(\ref{seriesmathbbV}) 
\beq
\la{\ee^{-\mathbb V}}\rao = \la\ee^{-{\mathbb V}_1}\rao 
- n^{-\half} \la{\mathbb V}_2\,\ee^{-{\mathbb V}_1}\rao 
+ \ldots\,.
\label{expmathbbV}
\eeq
Substitution of (\ref{expmathbbV}) in (\ref{splitlogpn}) yields
for $\log p_n$ the expansion
\beq
\log p_n=\log p_n^{(0)}+\log\la\ee^{-{\mathbb V_1}}\rao
- n^{-\frac{1}{2}} \la{\mathbb V}_2\rae
+\ldots,
\label{pertserlogpn}
\eeq
where for any function $F$ of the angular variables we define
the {\it first order average}
\beq
\la F\rae = 
\frac{\la F\,\ee^{-{\mathbb V}_1}\rao}{\la\ee^{-{\mathbb V}_1}\rao}.
\label{averages1F}
\eeq
The full average of a function $F$ is given by
\beq
\la F\ra = \la F\rae -n^{-\half} [\,\la F{\mathbb V}_2 \rae
- \la F\rae \la{\mathbb V}_2\rae\,] +\ldots\,. 
\label{pertserF}
\eeq
If we can substantiate
our claim that ${\mathbb V}_1$ is of order $n^0$  
and $n^{-\half}{\mathbb V}_2$ of higher order, then
(\ref{pertserlogpn}) and (\ref{pertserF})
are perturbation series in $n^{-\half}$. 

In this section we study ${\mathbb V}$
with the sole purpose of isolating an appropriate expression ${\mathbb V}_1$.
This will require an important amount of work.
In section \ref{secV} we will show the anticipated result, {\it viz.}
that for $n\to\infty$ the average 
$\la\ee^{-{\mathbb V}_1}\rao$ tends to a finite constant $C$.
In view of equation (\ref{pertserlogpn})
this then establishes the asymptotic expansion of $\log p_n$ up to terms that
vanish as $n\to\infty$.
In sections \ref{secV} and \ref{secdeviations} we will determine first 
order expressions for various correlation functions.


\subsection{An analog: the theory of elasticity}
\label{secstrategy}

Our determination of 
the interaction ${\mathbb V}_1$ will be based on the following
consideration. 
The regular $n$-sided polygon
is an obvious point of symmetry in phase space.
In fact, Drouffe and Itzykson \cite{DI84} 
constructed an exact lower bound for $p_n$ by expanding
the vertex vectors $\bS_m$ about their regular polygon positions.
Such an expansion preserves, however, 
the long range order of the regular polygon and
is therefore likely to cover only a small fraction of the effectively
contributing phase space.
Therefore we will proceed here as in the theory of
elasticity, where one considers the deviations 
not of the {\it positions,} but of the {\it interatomic
distances,} from their regular lattice values.
The analogs of the interatomic distances are the angles
$\xi_m$ between the midpoint vectors and 
$\eta_m$ between the vertex vectors.
Hence we will expand ${\mathbb V}$ in powers of
$\delta x_m$ and $\delta y_m$ defined in (\ref{defdeltaxdeltaym}). 
This procedure allows, in principle, 
for large deviations from the regular $n$-sided polygon.
In the nontrivial and rather lengthy analysis 
of this section we will step by step identify the order in $n$
of the terms that we encounter, 
relegate to ${\mathbb V}_2$ those
of order $n^{-\half}$ or smaller, and retain the others.
The outcome of this will be expression 
(\ref{exprmathbbVFT}) for ${\mathbb V}_1$.
The terms that contribute to ${\mathbb V}_1$ are all
quadratic in the $\delta x_m$ and $\delta y_m$.


\subsection{The interaction ${\mathbb V}$ as a series in the angles}
\label{sermathbbV}

Definition (\ref{defmathbbV}) of 
$\exp(-{\mathbb V})$ is a product of factors
$G'$, $\rho_m$, $T^{\phantom{0}}_m\xi^{-1}_m$, and $1+n^{-1}V$.
We will successively carry out the analysis of the orders in $n$ 
appearing in each of these factors.
Conceptually two different steps must be distinguished, 
even though in practice we have not strictly separated them.
First, since the angles $\beta_m, \gamma_m, \xi_m,$ and $\eta_m$
are all small, even though not all of the same
order, it will be convenient
to expand all relevant expressions as power series in these
angles. This is a matter of practical convenience rather than of
principle. Second, we will convert the resulting series to an expansion 
in the $\delta x_m$ and $\delta y_m$, retaining only the leading 
order.


\subsubsection{The factors $G'$, $\rho_m\,$, 
and $\,T^{}_m\xi^{-1}_m$}
\label{secfactorsGrhoTxi}

We first consider the 
function $G$. It is given by (\ref{deff}) 
in terms of the $\beta_\ell$ and $\gamma_\ell$ which,
in turn, are expressed by (\ref{inversebgxy}) 
in terms of $\beta_1$  and of the $\xi_m$ and $\eta_m$.
When differentiating $G$ with
respect to $\beta_1$ at fixed $\xi$ and $\eta$ we obtain
\bea
G'(\xi,\eta;\beta_1) &=& \frac{1}{2\pi}\summ\big(
\tan\beta_m + \tan\gamma_m \big)\nonumber\\[2mm]
&=& 1 + \frac{1}{6\pi}\summ(\beta_m^3+\gamma_m^3)
+ \ldots\,,
\label{serGprime}
\eea
where in the second step we used the geometrical
sum rule (\ref{sumrulebetagamma}) 
to cancel the linear terms in the angles, and where
the dots stand for terms of fifth and higher order in 
$\beta_m$ and $\gamma_m$.
According to (\ref{b3plusc3}) the sum $\beta_m^3+\gamma_m^3$ 
scales as $\sim n^{-2}$. Since the sum on $m$
contributes an extra factor $n$,
we deduce from (\ref{serGprime}) that 
\beq
G'(\xi,\eta;\beta_*)=1+{\cal O}(n^{-1}),
\label{resGprime}
\eeq
{\it i.e.,} this derivative is to leading order
independent of its arguments. The 
${\cal O}(n^{-1})$ corrections in (\ref{resGprime}) may be treated
perturbatively ({\it i.e.,} they belong to ${\mathbb V}_2$),
but will not be dealt with in the present work.

Secondly, we obtain from (\ref{deftau}) and 
(\ref{sumruletau}) the expansion
\beq
\prod_{m=1}^n \rho_m^2=\exp \big(-\sum_{m=1}^n\tau_m^2 +
\tfrac{2}{3}\sum_{m=1}^n\tau_m^3 + \ldots \big).
\label{expprodrho}
\eeq
Expanding (\ref{condcosbcosg}) and 
using (\ref{deftau})
yields for $\tau_m$ the series
\beq
\tau_m = \tau_n +
\tfrac{1}{2} \sum_{\ell=1}^m(\beta_\ell^2-\gamma_\ell^2)
- \tfrac{1}{24} \sum_{\ell=1}^m(\beta_\ell^4-\gamma_\ell^4) + \ldots\,,
\label{reltaubc}
\eeq
which is valid for $m=1,2,\ldots,n$; 
we recall that for $m=n$ it is 
equivalent to the no-spiral constraint.

Equation (\ref{b4minc4}) tells us that 
$\beta_\ell^2-\gamma_\ell^2$ 
and $\beta_\ell^4-\gamma_\ell^4$ scale as $\sim n^{-\frac{3}{2}}$ 
and $n^{-\frac{5}{2}}$, respectively. 
When $m$ is of the order of $n$, as is typically the case, the sum on
$m$ contributes a factor $n$ to the scaling, so that $\tau_m-\tau_n$
is to leading order proportional to $n^{-\half}$. 
This leads us to define the auxiliary scaled variables $r_m$ by
\beq
\tau_m=n^{-\half}r_m\,, \qquad m=1,\ldots,n,
\label{defrm}
\eeq
which allow us to write the gauge condition as 
\beq
\summ r_m=0.
\label{sumruler}
\eeq
From (\ref{defrhom}), 
(\ref{deftau}), (\ref{defrm}), and (\ref{defrav})
we find that $r_m$ is related to the original midpoint distance $R_m$ by
\bea
R_m &=& R_{\rm av}(1+n^{-\half}r_m) \nonumber\\
&=& R_c + \pi^{-\half}(r_{\rm av}+r_m)
+ {\cal O}(n^{-\half}). 
\label{relRmrm}
\eea 
In terms of scaled variables (\ref{reltaubc}) becomes,
if we use both equations (\ref{b4minc4}),
\beq
r_m = r_n\,+\,\frac{1}{2n}\sum_{\ell=1}^m x_\ell (b_\ell-c_\ell) 
\,+\,\frac{1}{12n^2}\sum_{\ell=1}^m x_\ell b_\ell c_\ell (b_\ell-c_\ell)
\,+\,\dots\,,
\label{exprtaumsc}
\eeq
valid for $m=1,2,\ldots,n$, and where the dots stand for higher order terms. 
Let us now consider (\ref{expprodrho}).
Since $\tau_m^2 \sim n^{-1}$ and since
the sum on $n$ contributes an extra factor $n$, the first term
in the exponent in (\ref{expprodrho}) is of order $n^0$. We will keep
it. However, $\tau_m^3$ is of order $n^{-\frac{3}{2}}$,
so that its sum on $m$
will vanish for $n\to\infty$ and can be treated perturbatively. 
Hence the contribution to $\exp(-{\mathbb V})$ is a factor
\beq
\prod_{m=1}^n \rho_m^2=
\Big[ \exp\big(-\sum_{m=1}^n\tau_m^2 \big) \Big]\,
\big(1\,+\,{\cal O}(n^{-1})\big).
\label{resprodrhog}
\eeq

Thirdly we consider the product of the
$T_m^{\phantom{2}}\xi_m^{-1}$. Symmetrizing in $\beta_m$ and $\gamma_m$ 
with the aid of the no-spiral constraint $G=0$ we may write this product as
\bea
\prod_{m=1}^n T_m^{\phantom{2}}\xi_m^{-1} &=& 
\prod_{m=1}^n \frac{\sin(\beta_m+\gamma_m)}
{(\beta_m+\gamma_m)\cos\beta_m\cos\gamma_m}\nonumber\\
&=& \exp \sum_{m=1}^n 
\big( \tfrac{1}{2} (\beta_m^2+\gamma_m^2) 
- \tfrac{1}{3} (\beta_m+\gamma_m)^2 +\ldots \big). 
\label{expxiinvT}
\eea
The sum $\,\sum_m(\beta_m^2 + \gamma_m^2)\,$
contains $n$ terms of order $(n^{-\half})^2$,
is therefore itself of order $n^0$ and must be kept.
Since $\beta_m+\gamma_m=\xi_m$ scales as $n^{-1}$, the sum 
$\sum_m(\beta_m+\gamma_m)^2$ also scales as $n^{-1}$ and may be treated
perturbatively.
The higher order terms in the sum in the exponential in
(\ref{expxiinvT}) are at least quartic in the angles and may 
again be treated by perturbation theory. Hence, to leading order, 
\beq
\prod_{m=1}^n T_m^{\phantom{2}}\xi_m^{-1} =
\Big[ \exp \tfrac{1}{2} \sum_{m=1}^n (\beta_m^2+\gamma_m^2) \Big]\,
\big(1\,+\,{\cal O}(n^{-1})\big) 
\label{resexpxiinvT}
\eeq
is the contribution of this product to $\exp(-{\mathbb V})$.


\subsubsection{The factor $\,(1+n^{-1}V)^{-n}$} 
\label{secfactor1pV}

When writing (\ref{defV}) as we did, we anticipated
that $V$ thus defined would scale as $\sim n^0$ for
$n\to\infty$. 
We will show here that this is indeed the case. We expand 
\beq
(1+n^{-1}V)^{-n} =
\exp\big(-V+\tfrac{1}{2}n^{-1}V^2 +\ldots\big)
\label{expVn}
\eeq
and next wish to find $V$ as a series in the angles.
The decomposition (\ref{defV}) of ${\cal A}$ 
was motivated by the fact that in this way, and since
the $\tau_m$ are quadratic in the angles,
the series for $V$ contains terms
only from third order up. 
We expand $V$ as given by (\ref{exprVbetagammatau}) and find
\bea
V &=& \frac{n}{\pi} \summ (1+\tau_m)^2 
    \big[ \tfrac{1}{3}(\gamma_{m}^3 + \beta_{m+1}^3) 
     + \ldots \big] \nonumber\\
&& + \,\frac{n}{\pi} \summ \tau_m (\gamma_m + \beta_{m+1}) 
+ \frac{n}{2\pi} \summ \tau_m^2 (\gamma_m + \beta_{m+1}). 
\label{serVbetagammatau}
\eea
In the first line of (\ref{serVbetagammatau})
the dominant term is the one where 
\,$\frac{1}{3}(\gamma_{m}^3+\beta_{m+1}^3)$\,
is summed without any factor $\tau_m$. 
To estimate its order in $n$ 
we may invoke (\ref{c3plusb3}); we will however perform a slightly
refined analysis by using that 
$\gamma_m+\beta_{m+1}=\eta_m$, setting
as before
$\eta_m=\bar{\eta}+\delta\eta_m$ with $\bar{\eta}=2\pi n^{-1}$,
and writing
\bea
\gamma_{m}^3+\beta_{m+1}^3 &=&
  \tfrac{3}{2}\eta_m (\gamma_m^2 + \beta_{m+1}^2) 
- \tfrac{1}{2}\eta_m^3 \nonumber\\[2mm]
&=& \tfrac{3}{2}\bar{\eta}(\gamma_m^2 + \beta_{m+1}^2) 
+ \tfrac{3}{2}\delta\eta_{m}(\gamma_m^2 + \beta_{m+1}^2) 
- \tfrac{1}{2}\eta_m^3.
\label{transfbg3}
\eea
Because of the scaling (\ref{scbetagamma}) we have 
$\bar{\eta}(\gamma_m^2 + \beta_{m+1}^2) \sim n^{-2}$. 
Hence, taking into account the effect 
of $n$ times the sum on $m$ in (\ref{serVbetagammatau}),
we conclude that the first term on
the RHS of (\ref{transfbg3}) yields a contribution of order $n^0$
to $V$ and should therefore be kept.
The term $-\half\eta_m^3$ in (\ref{transfbg3})
contributes to order $n^{-1}$ to $V$
and can be treated perturbatively.
The term $\tfrac{3}{2}\delta\eta_{m}(\gamma_m^2 + \beta_{m+1}^2)$
on the RHS of (\ref{transfbg3}) requires more attention. 
The factor $\delta\eta_m$ is of the same order as $\bar{\eta}$
in the first term, {\it viz.} $\sim n^{-1}$. 
However, $\delta\eta_m$ is a zero average random variable and 
therefore \cite{footnotecorr}
the sum on $m$ will raise the order by only $n^\half$ instead of $n$,
whence it follows that the resulting contribution to $V$
can be treated perturbatively.
Putting these arguments together we get for the first line of
(\ref{serVbetagammatau}) 
\beq
\frac{n}{\pi} \summ (1+\tau_m)^2 
\big[ \tfrac{1}{3}(\gamma_{m}^3 + \beta_{m+1}^3) + \ldots \big]\,=\,
\summ (\gamma_m^2+\beta_{m+1}^2)
\,+\,{\cal O}(n^{-\half}).
\label{thirdsum}
\eeq
 
In the second line of equation (\ref{serVbetagammatau}) 
we perform a similar analysis on the first sum. Using the sum rule
(\ref{sumruletau}) we can write
\beq
\frac{n}{\pi}\summ \tau_m (\gamma_m + \beta_{m+1}) = 
\frac{n}{\pi}\summ\tau_m\delta\eta_m\,. 
\label{firstsum}
\eeq
Since $\delta\eta_m\sim\eta_m\sim n^{-1}$, the product
$\delta\eta_m\tau_m$ is of order $n^{-\half}$, and when
summed on $m$ with the prefactor $n/\pi$ in
(\ref{serVbetagammatau}),
it would appear that it contributes a term of order $n^{\half}$ to $V$. 
This again overestimates the order. 
Since the $\delta\eta_m$ are zero average random variables,
we expect \cite{footnotecorr} in fact
that the sum will raise the order by
a factor of only $n^{\half}$, 
and hence that this term produces an order $n^0$ contribution to $V$.
Therefore we keep this term.

In the second sum in the second
line of (\ref{serVbetagammatau}) we write
$\tau_m^2 (\gamma_m+\beta_{m+1})
= 2\pi n^{-1}\tau_m^2 + \tau_m^2\delta\eta_m$. 
Both these terms are
of order $(n^{-\half})^2 \times n^{-1}=n^{-2}$ but the summation on $m$
increases the order of the first one by $n$ and of the second one, which is
random \cite{footnotecorr}, by only $n^{\half}$. 
Hence only the first term leads to a
contribution of order $n^0$ to $V$ and we get
\beq
\frac{n}{2\pi}\summ \tau_m^2 (\gamma_m+\beta_{m+1}) 
= \summ \tau_m^2 + {\cal O}(n^{-\half}).
\label{secondsum}
\eeq

Finally, we must consider
higher orders in $V$ in the expansion (\ref{expVn}).
Since we concluded that $V$ is of order $n^0$, it follows that
all higher order terms $n^{-1}V^2$, $n^{-2}V^3,$ $\ldots$ 
in the expansion of the logarithm 
can be treated perturbatively. Hence the contributions to $(1+n^{-1}V)^{-n}$ 
that must be kept are given by (\ref{thirdsum}), (\ref{firstsum}), 
and (\ref{secondsum}). Upon adding them together we are led to
\bea
(1+n^{-1}V)^{-n} &=& 
\exp \Big(\!- \summ\big(\gamma_m^2+\beta_{m+1}^2\big) - \summ\tau_m^2
- \frac{n}{\pi}\summ\tau_m\delta\eta_m \Big) 
\nonumber\\[2mm]
&& \times \big( 1\,+\,{\cal O}(n^{-\half}) \big),
\label{resexpVn}
\eea
which completes our discussion of the factors 
composing $\exp(-{\mathbb V})$.


\subsubsection{ Intermediate result for ${\mathbb V}$ }
\label{secintermediate}

Upon collecting the contributions identified above
in equations (\ref{resGprime}), (\ref{resprodrhog}),
(\ref{resexpxiinvT}), and (\ref{resexpVn}) we find 
\bea
{\mathbb V} &=& \tfrac{1}{2}\summ (\beta_m^2 + \gamma_m^2)
+ 2\summ \tau_m^2 + \frac{n}{\pi}\summ \tau_m \delta\eta_m 
\nonumber\\
&=& \frac{1}{n}\summ \big( \tfrac{1}{2} (b_m^2+c_m^2) + 2r_m^2 \big) 
\,+\, \frac{1}{\pi n^{\frac{1}{2}}} \summ r_m \delta y_m
\,+\,{\cal O}(n^{-\half}),\phantom{xx} 
\label{defmathbbV0}
\eea
where in the second line we have passed to scaled variables.
Although this expression has the appearance of a quadratic form,
we recall that $r_m$ is itself 
a series in the angles starting out with a quadratic term. 

The important facts established at this point are 
(i) that ${\mathbb V}$ given by (\ref{defmathbbV0})
is of order $n^0$ as $n\to\infty$; and (ii) that the terms neglected 
in (\ref{defmathbbV0}) may be classified by their order in $n^{-\half}$ 
and can hence be arranged, if desired, in a perturbation series. 
\vspace{2mm}

A word of caution is in place concerning the averages $\la F \ra$.
To arrive at ${\mathbb V}$ of equation (\ref{defmathbbV0}), 
we have used twice in
section \ref{secfactor1pV} that the sum of $n$ sufficiently weakly correlated
random variables has a random part which is 
of relative order order $n^{-\half}$ with respect to its systematic part.
For a two-point correlation
at a distance $k$ (measured in units of the midpoint index $m$) 
we expect that in an analogous 
calculation sums of $k$ terms will
intervene, for which the random part is of relative 
order $k^{-\half}$ with respect to the systematic part. 
Hence the resulting expression (\ref{defmathbbV0}) can be used for 
correlation functions at distances $k \sim n$, but
in the absence of further analysis {\it not} for $k \sim n^0$.
We speculate that in fact it is valid for correlations
at all distances $k \sim n^\epsilon$ for any $\epsilon>0$,
but our further work will not depend on that.


\subsection{Expressing ${\mathbb V}$ in terms of $\delta x_m$ and
$\delta y_m$}
\label{secV0dxdy}

We return to the study of expression (\ref{defmathbbV0}) 
for ${\mathbb V}$. In order to prepare for integrating 
$\exp(-{\mathbb V})$ on the $x_m$ and $y_m$,
as required by the average $\la\exp(-{\mathbb V})\rao$
in (\ref{pertserlogpn}), we first wish to express
the $b_m$, $c_m$, and $r_m$ that appear in (\ref{defmathbbV0}) 
in terms of the $\delta x_m$ and $\delta y_m$.


\subsubsection{Recursion relation}
\label{secrecursionrel}

In principle, this should be done as follows. 
We may deduce from (\ref{exprtaumsc}) the recursion relation
\beq
r_m - r_{m-1} = n^{-1} x_m u_m + {\cal O}(n^{-2}), \qquad
m=1,\ldots,n,
\label{recr}
\eeq
with $r_0=r_n$ as an additional condition expressing the no-spiral
constraint and with the $u_m$ defined by
\beq
u_m = \tfrac{1}{2} (b_m-c_m), \qquad m=1,\ldots,n.
\label{defu}
\eeq
From (\ref{defu}) and (\ref{inversebgxysc}) it follows that the $u_m$
satisfy
\beq
u_m - u_{m-1} = n^{-\half} f_{m-1}\,, \qquad m=2,\ldots,n,
\label{recu}
\eeq
in which the $f_m$ are defined as
\beq
f_{m-1} = -\tfrac{1}{2} (\delta x_m + \delta x_{m-1} -2\delta y_{m-1}),
\qquad m=1,\ldots,n
\label{deffm}
\eeq
with the conventions $\delta x_0=\delta x_n$,\, $\delta y_0=\delta y_n$, and
$f_0=f_n$. The extension of (\ref{recu}) to $m=1$ defines $u_0$.
Because of the geometrical sum rules (\ref{sumrulexi}) and 
(\ref{sumruleeta}) we have $u_0=u_n$ as well as
\beq
\summ f_m=0.
\label{sumrulef}
\eeq
If initial values $r_0$ and $u_0$ are given,
the system of equations (\ref{recr}) and (\ref{recu}) 
may be solved recursively in the midpoint index $m$,
yielding the $r_m$ and $u_m$ in terms of the $f_m$, and hence in terms 
of the $x_m$ and $y_m$.
In the end the initial values are determined by two conditions, {\it viz.} 
the no-spiral constraint $r_n=r_0$ and the gauge condition $\summ r_m=0$.
Proceeding this way gives a host of terms
many of which contain multiple sums on the midpoint index.
Their order in $n$ may be determined by the methods 
of section \ref{sermathbbV}. This route, however, is cumbersome as too much
detailed information is carried along.
We will therefore adopt a different but
equivalent approach which selects the leading order terms
more easily.
It also reestablishes the cyclic invariance in the index $m$ and the symmetry
under reversal of orientation. 

In (\ref{recr}) we substitute $x_m=\bar{x}+\delta x_m$.
Of course $\delta x_m$ is of the same order as $\bar{x}$.
But since the $\delta x_m$ are independent zero average random variables,
by a reasoning analogous to the one advanced in subsection 
(\ref{secfactor1pV}), 
the random part, when summed on $m$ 
[as for exemple in (\ref{exprtaumsc})], 
is of order $n^{-\half}$ relative to the systematic part.
It may therefore be neglected to lowest order.
Upon combining (\ref{recr}) and (\ref{recu}) and setting
$\bar{x}=2\pi$ we then find for $r_m$ the second order recursion
\beq
r_{m+1}-2r_m+r_{m-1}= 2\pi n^{-\frac{3}{2}} f_{m}\,,
\label{rec2r}
\eeq
which holds the key to much of what follows.


\subsubsection{Solution in terms of Fourier transforms}
\label{secsolutionFT}

In order to solve the recursion (\ref{rec2r}) 
we define the Fourier transforms
\beq
\hat{r}_q=n^{-1}\sum_{m=1}^n \ee^{2\pi{\rm i} qm/n} r_m\,,
\qquad
\hat{u}_q={n}^{-1}\sum_{m=1}^n 
\ee^{2\pi{\rm i} q(m-\frac{1}{2})/n} u_m\,,
\label{defrquq}
\eeq
and 
\beq
\hat{f}_q= \frac{1}{2\pi n^\half}\summ \ee^{2\pi{\rm i} qm/n} f_m\,,
\label{defFq}
\eeq 
where $q=0,\pm 1,\pm 2,\ldots,\pm(\frac{1}{2}n-\frac{1}{2})$ 
for $n$ odd and 
$q=0,\pm 1,\pm 2,\ldots,\pm(\frac{1}{2}n-1),\frac{1}{2}n$ 
for $n$ even.
Here $q$ is the `wavenumber' and $2\pi/q$ the
`wavelength' along the cell perimeter, the scaled index $2\pi m/n$ 
being equal to the polar angle in the plane up to corrections of
order $n^{-\half}$.
In Fourier language the unknowns $u_m$ and $r_m$
of (\ref{recu}) and (\ref{rec2r}) are readily solved 
in terms of the forces $f_m$
and one finds
\bea
\hat{u}_q &=& -\frac{\pi}{ {\rm i} n \sin\frac{\pi q}{n} }
\hat{f}_q
\,\,\simeq\,\, -\frac{1}{{\rm i}q}\hat{f}_q\,,
\qquad q \neq 0,
\label{soluq}\\[2mm]
\hat{r}_q &=& -\frac{\pi^2}{ n^2 \sin^2\frac{\pi q}{n} } \hat{f}_q\
\,\,\simeq\,\, -\frac{1}{q^2}\hat{f}_q\,,
\qquad q \neq 0.
\label{solrq}
\eea
The no-spiral constraint and the gauge condition become
$\hat{u}_0=0$ and $\hat{r}_0=0$, respectively.
The $\simeq$ signs in (\ref{soluq}) and (\ref{solrq}) 
indicate the limit $n\to\infty$ at fixed $q$
and amount to neglecting higher orders in $q/n$;
these are small if we restrict ourselves to $q$ values
on a scale $\sim n^{1-\epsilon}$,
{\it i.e.} to spatial distances which, 
in units of the midpoint index $m$,
scale as $\sim n^\epsilon$ with $\epsilon>0$.

Next, Fourier transforms
of $\delta{x}_q$ and $\delta{y}_q$ will be defined as
\beq
\hat{X}_q= \frac{1}{2\pi n^\half}\summ \ee^{2\pi{\rm i} qm/n} \delta x_m\,,
\qquad 
\hat{Y}_q= \frac{1}{2\pi n^\half}\summ \ee^{2\pi{\rm i} qm/n} \delta y_m\,. 
\label{defXqYq}
\eeq
From (\ref{deffm}) and (\ref{defXqYq}) it then follows that
\beq
\hat{f}_q \,\,=\,\, -\tfrac{1}{2} (1+\ee^{-2\pi{\rm i}q/n}) \hat{X}_q 
+ \hat{Y}_q \,\,\simeq\,\, -(\hat{X}_q-\hat{Y}_q).
\label{relhatFXY}
\eeq

We remark that the factor $n^{-\half}$ in (\ref{defXqYq})
is the usual one for a Fourier transform of $n$ independent variables;
it guarantees that $\hat{X}_q$ and $\hat{Y}_q$ are of order $n^0$.
The factor $n^{-1}$ in the Fourier transforms (\ref{defrquq}) and 
(\ref{defFq}) then follows from the requirement that in each of the equations 
(\ref{soluq}), (\ref{solrq}), and (\ref{defFq}) all members be of the same
order in $n$. 


\subsubsection{Defining ${\mathbb V}_1$}
\label{secdefV1}

Employing (\ref{defrquq}) and (\ref{defXqYq}) 
we are able to express ${\mathbb V}$ of equation
(\ref{defmathbbV0}) as 
\beq
{\mathbb V} = \sum_{q \neq 0}\,
[ \hat{u}_q\hat{u}_{-q} + 2\hat{r}_q\hat{r}_{-q} + 
2\hat{r}_q\hat{Y}_{-q} ] 
+ {\cal O}(n^{-\half}),
\label{precursorVFT}
\eeq
where $q$ runs through all nonzero integers and where
we Fourier transformed using that $b_m$ and $-c_m$ are both 
equal to $u_m+{\cal O}(n^{-\half})$. 
We now insert the solutions (\ref{soluq}) and (\ref{solrq}) in
(\ref{precursorVFT}). We relegate to the term 
$n^{-\half}{\mathbb V}_2 $ in (\ref{seriesmathbbV}) also the higher
orders in $q/n$ discussed above and arrive at
\beq
{\mathbb V} = {\mathbb V}_1 + {\cal O}(n^{-\half})
\eeq
where we finally decide to define ${\mathbb V}_1$ as
\beq
{\mathbb V}_1 = \sum_{q \neq 0}\,
\big[ \big(\frac{1}{q^2} + \frac{2}{q^4}\big)
(\hat{X}_{q}-\hat{Y}_{q})(\hat{X}_{-q}-\hat{Y}_{-q})
 + \frac{2}{q^2}(\hat{X}_{q}-\hat{Y}_{q})\hat{Y}_{-q} \big].
\phantom{xx}
\label{exprmathbbVFT}
\eeq
In (\ref{exprmathbbVFT}) the sum on $q$ converges due to the presence of the
factors $1/q^2$ and $1/q^4$.
This confirms again that ${\mathbb V}_1$ is of order $n^0$.
The contributions to ${\mathbb V}_1$ come essentially
from wavelengths at the scale of the cell itself
and are fully separated in scale from the independent
local degrees of freedom that entered the noninteracting calculation. 
\vspace{2mm}

In this section we have extracted from ${\mathbb V}$, defined by
(\ref{defmathbbV}),  
its leading order term ${\mathbb V}_1$, given by (\ref{exprmathbbVFT}).
With this expression at our disposal we can now go ahead and
calculate the consequences of the interaction.
Expression (\ref{exprmathbbVFT}) could be considered as an
elastic Hamiltonian in the harmonic approximation, 
if it were not for the fact that the $\hat{X}_q$ and
$\hat{Y}_q$ are non-Gaussian variables. It will appear below
that the deviations from Gaussian statistics affect the results of 
interest only to higher order in $n^{-\half}$. 


\section{First order}
\label{secV}

At the end of section \ref{secexprmathV} we concluded
that ${\mathbb V}$ is of order $n^0$. This means, more
precisely, that it is $\sim n^0$ in the
region of phase space anticipated to contribute
dominantly to the average $\la\exp(-{\mathbb V}_1)\rao$.
We therefore expect this average also to be of order $n^0$.
By showing that below we will accomplish an essential step.
The development of this and the following
section are standard mathematical physics.


\subsection{Calculation of $p_n$}
\label{seccalcpn}

We consider the 
quadratic form ${\mathbb V}_1$ given by (\ref{exprmathbbVFT}) above.
It is useful to set 
\beq
{\bsZ}_q=(\hat{X}_q,\hat{Y}_q)
\label{defZ}
\eeq
 and to
write (\ref{exprmathbbVFT}) more compactly as
\beq
{\mathbb V}_1=\sum_{q\neq 0}\, 
{\bsZ}_q\cdottt{\bsV}_q\cdottt{\bsZ}_{-q}^{\rm T}\,,
\label{V0compact}
\eeq
where the superscript 
${\rm T}$ indicates transposition and where
\beq
{\bsV}_q = \left(
\begin{array}{rr}
q^{-2}+2q^{-4}\phantom{xx} & -2q^{-4} \\[2mm]
-2q^{-4}        \phantom{xx} & -q^{-2}+2q^{-4}
\end{array}
\right).
\label{exprBq}
\eeq
The exponential $\exp(-{\mathbb V}_1)$ with ${\mathbb V}_1$ given by
(\ref{V0compact}) has to be 
substituted for $\exp[-{\mathbb V}(x,y)]$ in (\ref{LTinvscalinguv}).
The variables of integration $x_m$ and $y_m$ are now coupled 
by ${\mathbb V}_1$. In order to decouple them we
apply a Hubbard-Stratonovich transformation.
We write
\beq
\bspsi_q=(\hat{\psi}^x_q,\hat{\psi}^y_q)
\label{defphipsi}
\eeq
and represent $\exp(-{\mathbb V}_1)$ as the integral 
\beq
\ee^{-{\mathbb V}_1} =
\int \Big[ \prod_{q\neq 0} \dd{\bspsi}_q\,
\frac{(\det\bsV_q^{-1})^\half}{4\pi}
\Big]\,\exp\Big[\! - \tfrac{1}{4}\sumqno 
{\bspsi}_q\cdottt\bsV_q^{-1}\cdottt\bspsi^{\rm T}_{-q} 
+ {\rm i} \sumqno\bspsi_q\cdottt\bsZ_{-q}^{\rm T} \Big]. 
\label{expV0intrepr}
\eeq
We have that $\,\det\bsV_q=-q^{-4}\,$ is negative and the paths
of integration of the $\hat{\psi}^x_q$ and $\hat{\psi}^y_q$
in the complex plane should be chosen such 
that the integrals in (\ref{expV0intrepr}) exists.
Let  $\psi^x_m$ and $\psi^y_m$ be related to $\hat{\psi}^x_q$ and
$\hat{\psi}^y_q$ in the same way that $\delta x_m$ and $\delta y_m$ are
related to $\hat{X}_q$ and $\hat{Y}_q$,
{\it i.e.} by (\ref{defXqYq}).
We set $\hat{\psi}^x_0=\hat{\psi}^y_0=0$. Then
\bea
\sum_{q\neq 0}\, \bspsi_q\cdottt\bsZ_{-q}^{\rm T}
&=&\sumqno\, (\hat{\psi}^x_q\hat{X}_{-q} + \hat{\psi}^y_q\hat{Y}_{-q})
\nonumber\\
&=&\frac{1}{(2\pi)^2}\summ (\psi^x_m\delta x_m + \psi^y_m\delta y_m).
\label{Par}
\eea
We insert (\ref{Par}) in (\ref{expV0intrepr})
and the integral representation (\ref{expV0intrepr})
in (\ref{LTinvscalinguv}).
We commute the $\bspsi_q$ integrations to the outside and are then able
to perform the integrals on the $x_m$ and $y_m$.
The result is 
\bea
p_n &\simeq& 
\int\Big[ \prod_{q\neq 0} \dd{\bspsi}_q\,
\frac{(\det\bsV_q^{-1})^\half}{4\pi}
\Big]
\exp\Big[ -\tfrac{1}{4}\sum_{q\neq 0} 
{\bspsi}_q\cdottt\bsV_q^{-1}\cdottt\bspsi_{-q}^{\rm T}\Big] 
\nonumber\\
&& \times\,\frac{(n-1)!}{2n}\pi^{-n}n^{-3n}\, 
\Big[\frac{n}{2\pi{\rm i}}\int\!\dd u\, \frac{\ee^{2\pi nu}}{u^{2n}}
\prod_{m=1}^n\Big(1-\frac{{\rm i}\psi^x_m}{4\pi^2 u}\Big)^{-2}\,\Big] 
\nonumber\\
&& \times \Big[ \frac{n}{2\pi{\rm i}}\int\!\dd v\,
\frac{\ee^{2\pi nv}}{v^n}
\prod_{m=1}^n\Big(1-\frac{{\rm i}\psi^y_m}{4\pi^2 v}\Big)^{-1}\,\Big].
\label{Pintuv}
\eea
Hence one sees that
the $u$ and $v$ integrals factorize again. The products on $m$
represent an insertion of the type $h(u,v)$ discussed in section 
\ref{secPnzerothorder}, which may be taken
at the saddle points $u=u_c=\pi^{-1}$ and $v=v_c=(2\pi)^{-1}$.
Doing so and exponentiating these products we get 
\bea
p_n &\simeq& p_n^{(0)} 
\int\Big[ \prod_{q\neq 0} \dd{\bspsi}_q\,
\frac{(\det\bsV_q^{-1})^\half}{4\pi}
\Big]
\exp\big[-\tfrac{1}{4}\sum_{q\neq 0} 
{\bspsi}_q\cdottt\bsV_q^{-1}\cdottt\bspsi^{\rm T}_{-q}\big] 
\nonumber\\
&& \times\exp\Big[
-\tfrac{1}{4}\sum_{q\neq 0}\hat{\psi}^x_q\hat{\psi}^x_{-q}
-\summ 2{\cal R}(\psi^x_m) \Big] \nonumber\\
&& \times\exp\Big[
-\tfrac{1}{2}\sum_{q\neq 0}\hat{\psi}^y_q\hat{\psi}^y_{-q}
-\summ {\cal R}(2\psi^y_m) \Big],
\label{Pintpsi}
\eea
where $p_n^{(0)}$ was given in (\ref{resP0}) and
the `remainder' ${\cal R}(x)$ stands for the infinite series
\beq
{\cal R}(x)=\sum_{k=3}^\infty 
\frac{1}{k} \Big(\frac{{\rm i}x}{4\pi}\Big)^k.
\label{defcalR}
\eeq
We will delay to section \ref{sectermscalR} the proof
that in the limit $n\to\infty$ the remainders in (\ref{Pintpsi})
contribute only to relative order $n^{-1}$.
Admitting this here and setting
\beq
\bsE=\left(
\begin{array}{rr}
1\phantom{xx} & 0 \\[1mm]
0\phantom{xx} & 2
\end{array}
\right)
\label{exprE}
\eeq
we therefore have 
\bea
p_n &\simeq& p_n^{(0)} 
\int\Big[ \prod_{q\neq 0} \dd{\bspsi}_q\,
\frac{(\det\bsV_q^{-1})^\half}{4\pi}
\Big]
\exp \Big[ -\tfrac{1}{4}\sum_{q\neq 0} 
{\bspsi}_q\cdottt(\bsE+\bsV_q^{-1})\cdottt\bspsi^{\rm T}_{-q} \Big] 
\nonumber\\
&=& p_n^{(0)} 
\prod_{q\neq 0} \frac{(\det\bsV_q^{-1})^\half}
      {\big[\det(\bsE+\bsV_q^{-1})\big]^\half} 
\nonumber\\[2mm]   
&=& p_n^{(0)}\prod_{q=1}^\infty\, \Lambda_q^{-1},
\label{exprCdet}
\eea
where we abbreviated
\bea
\Lambda_q &=& \det(\bse+\bsV_q\bsE) \nonumber\\[1mm] 
&=& 1-q^{-2}+4q^{-4}
\label{defmuq}
\eea
with $\bse$ standing for the $2\times 2$ unit matrix.
Upon combining (\ref{exprCdet}) and (\ref{defmuq}) 
with expression (\ref{resP0}) for 
$p_n^{(0)}$ one finds for $p_n$ the result announced in 
(\ref{resultpn})-(\ref{resultC}) of the introduction, 
\beq
p_n = \frac{C}{4\pi^2}\,
\frac{(8\pi^2)^n}{(2n)!} \big[\,1\,+\,{\cal O}(n^{-\half})\,\big],
\label{finalrespn}
\eeq
where
\beq
C = \prod_{q=1}^\infty \big(1-q^{-2}+4q^{-4}\big)^{-1}.
\label{finalresC}
\eeq
In the course of 
the calculations of the previous sections we have encountered  
correction terms of relative order $n^{-\half}$. 
In equation (\ref{finalrespn}) we have included these ${\cal O}(n^{-\half})$ 
terms again,
although strictly speaking we have not shown that their net effect does not
happen to vanish.

This is the first analytic determination of the large $n$ behavior of $p_n$.
Conjectures 
in earlier work (see \cite{Okabeetal00}) 
have almost invariably had the form of
decaying exponentials, whether or
not multiplied by a power of $n$, or sums of such expressions.
Only Drouffe and Itzykson \cite{DI84} 
come close to our analytic expression:
they assume that to leading order\, $\log p_n \!\simeq\! \alpha n$\,
and estimate $\alpha\approx -2$, a value confirmed by
(\ref{finalrespn}).
We refer to \cite{HJHletter05} for a plot of the asymptotic expression 
(\ref{finalrespn})
and a comparison with Drouffe and Itzykson's simulation data.


\subsection{Correlations to first order}
\label{seccorrelations}

The calculational technique that led to the expression for $p_n$ in the
preceding subsection, may also be used to find correlation
functions at long distances.


\subsubsection{Long distance correlations along the perimeter}
\label{seclongdistance}

We will study the correlations between the angles $\xi_\ell$ and
$\eta_m$ in terms of their scaled counterparts $x_\ell$ and $y_m$.
Passing to Fourier transforms 
with the aid of (\ref{defXqYq}) we have
\beq
\la \delta x_\ell\delta x_m\ra=\frac{(2\pi)^2}{n}
\sumqno\ee^{2\pi{\rm i}(k-\ell)q/n} \la \hat{X}_q\hat{X}_{-q} \ra
\label{corrFT}
\eeq
and analogous relations for $\la \delta y_\ell\delta y_m\ra$
and the cross correlation $\la \delta x_\ell\delta y_m\ra$.
From definition (\ref{averages1F}) and
equations (\ref{expV0intrepr}) and (\ref{Par})
it is clear that we obtain the first order expression
$\la \hat{X}_{q}\hat{X}_{-q} \rae$
by inserting in the integrand of (\ref{expV0intrepr}) 
a second derivative
$-\partial^2/\partial\hat{\psi}^x_q\partial\hat{\psi}^x_{-q}$
that acts on the factor 
$\exp(-{\rm i}\sumqno\bspsi_q\cdottt\bsZ_{-q}^{\rm T})$.
Carrying the calculation through we find that we have to differentiate
the exponential factors in the second 
line of (\ref{Pintpsi}), whence we have, for $q\neq 0$,
\beq
\la \hat{X}_q\hat{X}_{-q} \rae
=\tfrac{1}{2}-\tfrac{1}{4}\la\hat{\psi}^x_q\hat{\psi}^x_{-q}\raG
+\ldots\,.
\label{relcorrphipsiXX}
\eeq
Here $\la\ldots\raG$ is a Gaussian average with respect to the weight
${\rm cst}\times
\exp \big[ -\tfrac{1}{4}\sum_{q\neq 0} 
{\bspsi}_q\cdottt(\bsE+\bsV_q^{-1})\cdottt\bspsi^{\rm T}_{-q} \big] $
and the dots represent a well-defined
series of quartic and higher Gaussian averages
coming from the expansion of the remainder ${\cal R}$ in (\ref{Pintpsi}).
When evaluated, these higher order terms contain factors $n^{-1}$,
as will be shown in section \ref{sectermscalR}.
After completing the calculation for the other basic correlation functions
we obtain
\bea
\la \hat{X}_q\hat{X}_{-q} \rae
&\simeq& \tfrac{1}{2}-\tfrac{1}{4}\la\hat{\psi}^x_q\hat{\psi}^x_{-q}\raG\,,
\nonumber\\
\la \hat{Y}_q\hat{Y}_{-q} \rae 
&\simeq& 1\, - \,\la\hat{\psi}^y_q\hat{\psi}^y_{-q}\raG\,,
\nonumber\\
\la \hat{X}_q\hat{Y}_{-q} \rae
&\simeq& \phantom{1\,} - 
\,\tfrac{1}{2}\la\hat{\psi}^x_q\hat{\psi}^y_{-q}\raG\,,
\qquad q \neq 0,
\label{relcorrphipsiXY}
\eea
where the $\simeq$ sign indicates, as before, asymptotic equality
for $n\to\infty$. 
In view of (\ref{pertserF}) we may drop the index $1$ on the averages in 
(\ref{relcorrphipsiXY}).
The three Gaussian averages in (\ref{relcorrphipsiXY}) 
are due to the interaction ${\mathbb V}_1$. They come in addition to the
constants ($\half$, 1, and 0) that represent the
Fourier transformed noninteracting averages (\ref{exprcorrdeltaxdeltay}).
Straightforward calculation yields 
\bea
\la\hat{\psi}^x_q\hat{\psi}^x_{-q}\raG &=& 
\phantom{-}2\Lambda_q^{-1}q^{-2},
\nonumber\\
\la\hat{\psi}^y_q\hat{\psi}^y_{-q}\raG &=&
-2\Lambda_q^{-1}(q^{-2}-q^{-4}),
\nonumber\\
\la\hat{\psi}^x_q\hat{\psi}^y_{-q}\raG &=&
-4\Lambda_q^{-1}q^{-4},
\qquad q \neq 0,
\label{rescorrphipsi}
\eea
From (\ref{pertserF}), (\ref{rescorrphipsi}), and (\ref{relcorrphipsiXY})
we obtain for the basic correlation functions
the following set of first order expressions,
\bea
\la \hat{X}_q\hat{X}_{-q} \ra
&\simeq& \tfrac{1}{2} - \tfrac{1}{2}\Lambda_q^{-1}q^{-2},
\nonumber\\
\la \hat{Y}_q\hat{Y}_{-q} \ra 
&\simeq& 1 + 2\Lambda_q^{-1}(q^{-2}-q^{-4}),
\nonumber\\
\la \hat{X}_q\hat{Y}_{-q} \ra
&\simeq& \phantom{xxx} 2\Lambda_q^{-1} q^{-4},
\qquad q \neq 0,
\label{rescorrXY}
\eea
which represent the main result of this subsection.
The terms with the factor $\Lambda_q^{-1}$ in (\ref{rescorrXY})
have their origin in the interaction ${\mathbb V}_1$.
The important point about the $\hat{\psi}\hat{\psi}$
correlations is that they all decay at
least as $\sim 1/q^2$ when $q\to\infty$. This will turn out to
be an essential element in our analysis below.
Equations (\ref{rescorrXY}) may be
inverse Fourier transformed according to (\ref{corrFT}) 
and its analogs so as to yield expressions for
the spatial correlations
$\la \delta x_\ell\delta x_m\ra$, $\la \delta y_\ell\delta y_m\ra$, 
and $\la \delta x_\ell\delta y_m\ra$.
We will work these out in detail only for 
the more restricted correlation function 
$\la(\hat{X}_q-\hat{Y}_q)(\hat{X}_{-q}-\hat{Y}_{-q})\ra =
\la \hat{f}_q\hat{f}_{-q}\ra$
to be considered in section \ref{secdeviations}.
Because of our discussions on spatial scales 
at the end of section \ref{secintermediate} and following 
(\ref{soluq})-(\ref{solrq}),
the real space correlations so obtained are valid only
at distances large compared to the individual perimeter 
segments. 


\subsubsection{The remainders ${\cal R}$}
\label{sectermscalR}

Having gained control of the correlation functions, we can
now prove our claim that the remainders ${\cal R}$ 
in the exponentials in (\ref{Pintpsi}) 
contribute only to higher order in $n^{-1}$.
We may expand these exponentials as perturbation
series in the $\psi^x_m$ and $\psi^y_m$.
The odd terms vanish because of symmetry.
The lowest order contributing terms are therefore
proportional to $\sum_m(\psi^x_m)^4$ and $\sum_m(\psi^y_m)^4$.
Let us consider the latter.
It leads to a correction term which, relative to the term 
calculated, is of order $\sum_m\la(\psi^y_m)^4\raG$.
We therefore evaluate
\bea
\sum_m\la(\psi^y_m)^4\raG &=& \frac{(2\pi)^4}{n}
\!\sum_{q_1,q_2,q_3,q_4\neq 0} \!\delta_{q_1+q_2+q_3+q_4,0}\,
\la\hat{\psi}^y_{q_1}\hat{\psi}^y_{q_2}\hat{\psi}^y_{q_3}\hat{\psi}^y_{q_4}\raG
\nonumber\\[2mm]
&=& \frac{3(2\pi)^4}{n}
\Big[\sumqno\la\hat{\psi}^y_q\hat{\psi}^y_{-q}\raG\Big]^2,
\label{calcpsi4}
\eea
where we used Wick's theorem.
The fact that $\la\hat{\psi}^y_q\hat{\psi}^y_{-q}\raG$ decays as $q^{-2}$ is
crucial here. It means that the sum on $q$ inside the
brackets in (\ref{calcpsi4}) converges, and hence 
$\sum_m\la(\psi^y_m)^4\raG$
is of order $n^{-1}$. We therefore had the right to neglect the ${\cal R}$
terms in (\ref{Pintpsi}).
The other terms in the 
perturbation series in the $\psi^x_m$ and $\psi^y_m$
can be argued in a similar way to be at least of order $n^{-1}$.


\subsubsection{Debye-H\"uckel}
\label{secDebyeHuckel}

We remark that in order to arrive at 
the above results, as well as those that will follow in section
\ref{secdeviations},
one may bypass the Hubbard-Stratonovich transformation
by considering the variables of integration 
$\hat{X}_q$ and $\hat{Y}_q$ as zero mean Gaussian variables
of variances determined by those of
$\la\delta x_m^2\rao$ and $\la\delta y_m^2\rao$
given in (\ref{resavxy}), {\it i.e.} with weight
\beq
\exp\big(\! -\sumqno\bsZ_q\cdottt\bsE^{-1}\cdottt\bsZ_{-q} \big)
=\exp\Big(\! -\sumqno\hat{X}_q\hat{X}_{-q} 
          -\tfrac{1}{2}\sumqno\hat{Y}_q\hat{Y}_{-q} \Big).
\label{Gaussianweight}
\eeq
The procedure of replacing non-Gaussian variables by Gaussian ones
of the same mean and
variance is known in plasma physics as the Debye-H\"uckel
approximation; it is exact in the limit of high temperature.
This limit happens to apply to our case, where
the role of the temperature is played by $n$: the interaction `energy' 
${\mathbb V}_1$ would be extensive ({\it i.e.} proportional to $n$)
in an ordinary thermodynamical system, but is here of order
$n^0$, which is equivalent to having an inverse-temperature 
prefactor equal to $n^{-1}$. 

In the preceding subsection
we have chosen to go through the full Hubbard-Stratonovich
transformation. This has allowed us
to show that the procedure is exact
and to prepare the way for the calculation of higher order terms,
if desired, in the power series expansion in $n$.


\section{Weight functional for the perimeter}
\label{secdeviations}


\subsection{Weight function for the $\hat{f}_q$}
\label{secfphi}

Given that for large $n$ the perimeter approaches a circle,
it is natural to ask what the probability distribution is for 
its deviations from circularity.
This question has prompted work, in particular, 
in the mathematical literature \cite{Hugetal04,Calka02,CalkaSchreiber05},
but within the framework of large-cell limits different from ours. 
Deviations from circularity may be characterized in various ways,
{\it e.g.} as the difference between the radii of the smallest circumscribed
and the largest inscribed circle \cite{Hugetal04,CalkaSchreiber05}. 
Here we will exhibit in full generality
the weight of a specific trajectory of the perimeter,
characterized by the set of scaled `excess midpoint distances'
$\{r_1,r_2,\ldots,r_n\}$. The Fourier transforms $\hat{r}_q$ of these
distances are related to
the {\it differences\,} $\hat{X}_q-\hat{Y}_q\,$
by $\,q^2\hat{r}_q=-\hat{f}_q=\hat{X}_q-\hat{Y}_q$, 
which follows from (\ref{solrq}) and (\ref{relhatFXY}).
Hence one of the variables $\hat{X}_q$ and $\hat{Y}_q$ is redundant for
the description at which we are aiming.
The probability distribution of the $\hat{X}_q$ and $\hat{Y}_q$ is,
up to normalization, equal to the weight $\exp(-{\mathbb V}_1)$
with ${\mathbb V}_1$ given by (\ref{exprmathbbVFT}).
In order to derive from it the marginal distribution 
$P_1(\{ \hat{f}_q \})$ for the $\hat{f}_q$,
we short-circuit the Hubbard-Stratonovitch procedure
in the way indicated in section \ref{secDebyeHuckel}.
Upon integrating $\exp(-{\mathbb V}_1)$ directly
with the appropriate noninteracting
Gaussian weight (\ref{Gaussianweight}) we find 
the first order probability law
\beq
P_1(\{ \hat{f}_q \})
\simeq {\rm cst} \times
\exp\Big( \!-\tfrac{1}{3}\sumqno\Lambda_q \hat{f}_q\hat{f}_{-q} \Big)
\label{probdevf}
\eeq
with $\Lambda_q$ given by (\ref{defmuq}) and
where, because of (\ref{sumrulef}), it is understood that $\hat{f}_0=0$. 
The law (\ref{probdevf}) leads to the correlation
\beq
\la \hat{f}_q\hat{f}_{q'} \ra \simeq \tfrac{3}{2} \Lambda_q^{-1}
\delta_{q+q',0}\,,  \qquad  q,q' \neq 0,
\label{rescorrff}
\eeq
which may alternatively be derived 
from (\ref{rescorrXY}).
We remark parenthetically that
the weight (\ref{probdevf}) is suitable for computer simulation of
the large scale structure of Voronoi cells.
The rest of this section elaborates on equations (\ref{probdevf})
and (\ref{rescorrff}).


\subsection{Differential equation for $R(\phi)$}
\label{secdiffeqRphi}

For $n\to\infty$ it is possible on those scales
to consider the angle
\beq
\phi=\frac{2\pi m}{n}
\label{defphi}
\eeq
as a continuous variable \cite{footnoteangle} and to write
\beq
r_m=r(\phi),  \qquad f_m=\frac{2\pi}{n^\half}f(\phi).
\label{defrFphi}
\eeq
Recursion (\ref{rec2r}) 
relating the $r_m$ to the $f_m$ then becomes the second order differential
equation 
\beq
r''(\phi)=f(\phi).
\label{rec2rcont}
\eeq
Here $f(\phi)$ is Gaussian noise whose correlation function
$\la f(\phi)f(\phi') \ra$ we will now determine.
To this end we set $\phi'=2\pi m'n^{-1}$ and 
consider $\la f_m f_{m'} \ra$, 
which with the aid of (\ref{defFq}) may be written as
\bea
\la f_m f_{m'} \ra &=& \frac{(2\pi)^2}{n} \sum_q \sum_{q'}
\ee^{-2\pi{\rm i}(qm+q'm')/n} \la \hat{f}_q\hat{f}_{q'} \ra
\nonumber\\
&\simeq& \frac{3(2\pi)^2}{2n} \sum_{q \neq 0} 
\ee^{-{\rm i}q(\phi-\phi')} \Lambda_q^{-1},
\label{fmfmp}
\eea
where to arrive at the second line we used (\ref{rescorrff}).
Passing from $f_m$ to $f(\phi)$ 
and using (\ref{defmuq}) we obtain from (\ref{fmfmp})
\bea
\la f(\phi)f(\phi') \ra &\simeq& \tfrac{3}{2} 
\Big[ \sum_{q \neq 0} \ee^{-{\rm i}q(\phi-\phi')}\,
+\,\Gamma(\phi-\phi') \Big] \nonumber\\[2mm]
&=& \tfrac{3}{2}\big[ 2\pi\delta(\phi-\phi') - 1 + \Gamma(\phi-\phi') \big], 
\label{fphifphip}
\eea
where the $2\pi$-periodic function $\Gamma(\phi)$ is given by
\beq
\Gamma(\phi)= 2\sum_{q=1}^\infty
\frac{ q^{-2}-4q^{-4} }{ 1-q^{-2}+4q^{-4} }\,\cos q\phi.
\label{defGammaphi}
\eeq
This correlation function result merits a comment.
It is instructive to calculate, for comparison 
to (\ref{fmfmp}) and (\ref{fphifphip}),
also the simpler noninteracting average $\la f_mf_{m'} \rao$.
To this end one 
expresses $f_m$ and $f_{m'}$ in terms of the $\delta x_\ell$ and
$\delta y_\ell$ using (\ref{deffm}) and employs equations
(\ref{exprcorrdeltaxdeltay}). In the continuum limit it then appears
that in (\ref{fphifphip}) the terms
$2\pi\delta(\phi-\phi')-1$ represent the noninteracting result
and $\Gamma(\phi-\phi')$ the contribution generated by the interaction. 
The expression for this contribution was derived under the restrictive
condition that $|\phi-\phi'|\gg 2\pi n^{-1}$. We speculate that 
(\ref{fphifphip}) is valid, in
fact, for $|\phi-\phi'|$ scaling as $\sim n^{-1+\epsilon}$ with
arbitrary $\epsilon>0$.  
Finally we remark that expressions similar to (\ref{fphifphip}) 
for the correlations
$\la\delta x(\phi)\delta x(\phi')\ra$,
$\la\delta x(\phi)\delta y(\phi')\ra$, and
$\la\delta y(\phi)\delta y(\phi')\ra$
may be found in a fully analogous way.

In the continuum limit the Fourier transform of $\Lambda_q$ becomes
\beq
\sumqno \ee^{{\rm i}q(\phi-\phi')} \Lambda_q = 
2\pi\delta(\phi-\phi') - 1 + U(\phi-\phi')
\label{defU}
\eeq
with the $2\pi$-periodic function $U(\phi)$ given by
\beq
U(\phi) = -2\sum_{q=1}^\infty (q^{-2}-4q^{-4}) \cos q\phi. 
\label{exprUphisum}
\eeq
Similarly, in
(\ref{defU}) the term $U(\phi-\phi')$ represents the effect 
of the interaction. 
Both $\Gamma(\phi)$ and $U(\phi)$ are of zero average on the interval
$[0,2\pi]$.

\begin{figure}
\begin{center}
\scalebox{.50}
{\includegraphics{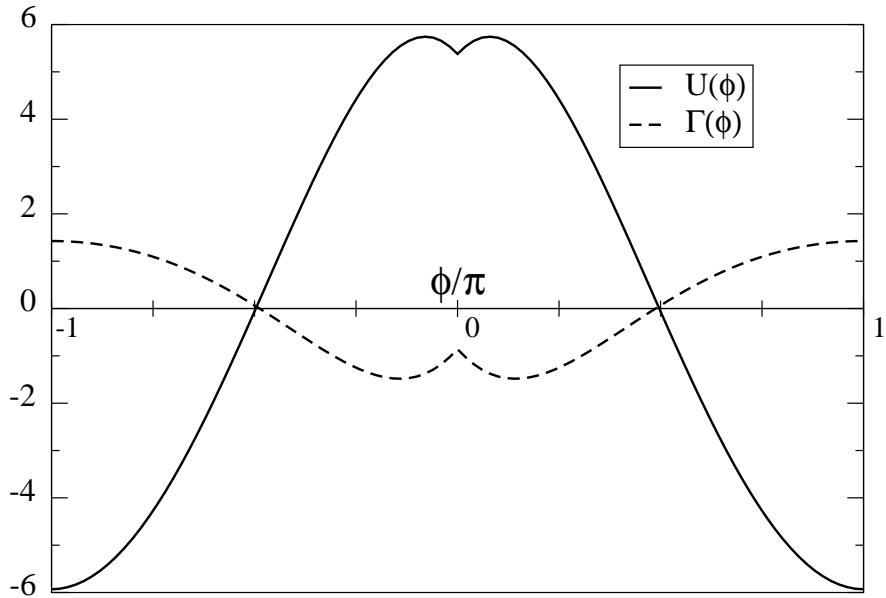}}
\end{center}
\caption{{\small The functions $U(\phi)$ and $\Gamma(\phi)$
of equations (\ref{exprUphi}) and (\ref{defGammaphi}), respectively.}}
\label{fig3}
\end{figure}

The sums on $q$ in equations (\ref{defGammaphi}) and (\ref{exprUphisum})
may be carried out analytically
[see reference \cite{GR}, equations (1.443.3), (1.443.6), and (1.445.2)].
One finds that (\ref{exprUphisum}) leads to the quartic polynomial
\beq
U(\phi) = -\tfrac{1}{6}\pi^2 \big( \tfrac{7}{15}\pi^2-1 \big)
+\tfrac{1}{2} \big( \tfrac{2}{3}\pi^2-1 \big) (\phi-\pi)^2
-\tfrac{1}{6}(\phi-\pi)^4, \quad 0\leq\phi\leq 2\pi,
\label{exprUphi}
\eeq
and its $2\pi$-periodic continuation. We have written it
such as to bring out its symmetry under the replacement 
$\phi\mapsto 2\pi-\phi$. 
The potential $U(\phi)$ in (\ref{exprUphisum}) consists of two terms
which merit separate discussion. 
The the first term is the one-dimensional electrostatic interaction 
but with opposite sign, so that it is attractive for like charges; 
and the second one is repulsive.
Near the origin we have the expansion
\beq
U(\phi) = U(0)+\pi|\phi|-(\tfrac{1}{2}+\tfrac{2}{3}\pi^2)\phi^2+\dots,
\label{Ulin}
\eeq
in which the linear term comes from the electrostatic interaction.
As a result
$U(\phi)$ has maxima at $\phi=\pm\phi_{_U}$,
where $\phi_{_U}=\pi[1-\{1-3/(2\pi^2)\}^\half]=0.079...\pi$.
We will not exhibit the analytic expression of $\Gamma(\phi)$ but have 
represented its graph in figure \ref{fig3}. 
It has minima at $\phi=\pm
\phi_{_\Gamma}$, where $\phi_{_\Gamma}\approx 0.139...\pi$.


\subsection{Weight functional for $R(\phi)$}
\label{secRphi}

Using (\ref{defU})
we subject the quadratic form in the exponential of (\ref{probdevf})
to an inverse Fourier transformation.
In the continuum limit
the probability distribution $P_1(\{ \hat{f}_q \})$ 
becomes a functional of $f(\phi)$,
or, because of (\ref{rec2rcont}), of $r''(\phi)$. We find
\bea
{P}_1[r(\phi)] &\simeq& {\rm cst}\times \exp\big(-{H}[r''(\phi)]\big),
\label{probdevr}\\[2mm]
{H}[r'' (\phi)] &=& \frac{1}{6\pi}
\int_0^{2\pi}\!\dd\phi\,[r''(\phi)]^2\nonumber\\[2mm]
&& 
+\,\frac{1}{12\pi^2} \int_0^{2\pi}\!\!\dd\phi_1 
\int_0^{2\pi}\!\!\dd\phi_2\,\, 
r''(\phi_1)U(\phi_1-\phi_2)r''(\phi_2).
\label{exprH}
\eea
Here the last line, which is the contribution from the interaction
${\mathbb V}$, is analogous to the Coulomb energy of a 
line charge density $r''(\phi)$ in a periodic one-dimensional system.
Equations (\ref{probdevr}) and (\ref{exprH}) are
valid for functions $r(\phi)$ coarse grained to a scale
much larger than the distance between the individual vertices, {\it i.e.,}
to angle differences $\Delta\phi \gg 2\pi n^{-1}$.
Moreover, the sum rules encountered during its derivation now lead to
three restrictions on the functions $r(\phi)$ that enter (\ref{exprH}):

(i) The geometrical sum rule (\ref{sumrulef}), taken in the continuum limit
and combined with the differential equation (\ref{rec2rcont}), implies
the `charge neutrality' condition 
$\int_0^{2\pi}\!\dd\phi\,r''(\phi)=r'(2\pi)-r'(0)=0$.
It follows that we may add an arbitrary constant to the potential $U$ in
(\ref{exprH}). 

(ii) The no-spiral constraint becomes
$\int_0^{2\pi}\!\dd\phi\,r'(\phi)=r(2\pi)-r(0)=0$.

(iii) The gauge condition takes the form
$\int_0^{2\pi}\!\dd\phi\,r(\phi)=0$.\\
\noindent This completes our discussion of the first order
probability functional $P_1$.

We now wish to reconvert the above findings
to results for the original midpoint distances 
$R_m=R(\phi)=R_c+\deltaR(\phi)$. 
From(\ref{relRmrm}) taken in the continuum limit and 
combined with (\ref{defdeltaR}) we find 
\beq
\deltaR(\phi) = \pi^{-\half} \big[ r_{\rm av} + r(\phi) \big] 
+ {\cal O}(n^{-\half}).
\label{relRphirphi}
\eeq
It follows that the probability distribution of $R(\phi)$ is determined by
the product of
$p_{\rm av}(r_{\rm av})$ given in (\ref{defprav}) and $P_1[r(\phi)]$ given in
(\ref{probdevr})-(\ref{exprH}).
Still setting
\beq
F(\phi) = \pi^{-\half}f(\phi) \qquad 
\label{tooldvar}
\eeq
we are led to the results (\ref{oderphi})-(\ref{integralcond}) 
announced in the introduction,
in which the factors $4\lambda$ have been restored.


\subsection{An application: fluctuations of $R_m$ around $R_{\rm av}$}
\label{secfluctuationsRm}

As an application we determine how, within a single realization of an
$n$-sided cell, the midpoint distances $R_m$ deviate from
their average $R_{\rm av}$.
This extends the considerations of section \ref{secdistributionRav}, where we
showed that $R_{\rm av}$ is itself within order $n^0$ from its ensemble
average $R_c=\pi^{-\half}n^\half$.

We have straightforwardly
\bea
n^{-1}\summ \la (R_m-R_{\rm av})^2\ra &=&
n^{-2}R_{\rm av}^2 \summ \la r_m^2\ra \nonumber\\
&=& n^{-1}R_{\rm av}^2 \sum_{q \neq 0} \la \hat{r}_q\hat{r}_{-q}\ra 
\nonumber\\
&=& n^{-1}R_{\rm av}^2 \sum_{q \neq 0} q^{-4} \la \hat{f}_q\hat{f}_{-q}\ra 
\nonumber\\ 
&=& \frac{3}{\pi}\sum_{q=1}^\infty q^{-4}(1-q^{-2}+4q^{-4})^{-1},
\label{resvarRm}
\eea
where the first step is based on the first line of (\ref{relRmrm}), in
the third step we used (\ref{solrq}), and in the last step we inserted
the explicit expressions for $R_{\rm av}$ and $\la\hat{f}_q\hat{f}_{-q}\ra$.
Equation (\ref{resvarRm}) yields the result announced in (\ref{varRm}).


\subsection{Convergence of $R(\phi)$ and $S(\phi)$}
\label{secRSphi}

The set of midpoints $\{\bR_m\,|\,m=1,\ldots,n\}$ and the set of vertices
$\{\bS_m\,|\,m=1,\ldots,n\}$ each define an $n$-sided polygon.
In the limit of large $n$
these polygons become the curves $R(\phi)$ and $S(\phi)$, respectively.
Of these, $S(\phi)$ is the true perimeter.
We show here that for $n\to\infty$ the curve 
$R(\phi)$ converges rapidly to
$S(\phi)$ so that in the discussion above
we were justified in referring to $R(\phi)$ 
as the perimeter.

Two points $\bR_m=(R_m,\Phi_m)$ and $\bS_m=(S_m,\Psi_m)$ 
have angular coordinates such that 
$|\Phi_m-\Psi_m|=|\gamma_m|={\cal O}(n^{-\half})$.
Since $\Phi_\ell$ increases by steps of order $\sim n^{-1}$ when $\ell$
increases by $1$, there exists a third point $\bR_{m'}=(R_{m'},\Phi_{m'})$,
with $m'$ depending on $m$, such that
\beq
|\Phi_{m'}-\Psi_m|={\cal O}(n^{-1}).
\label{diffPhipPsi}
\eeq
Typically we will have $|m'-m|={\cal O}(n^{\half})$.
In order to estimate how the 
distance between $\bR_{m'}$ and $\bS_m$ scales with $n$ we write
\beq
|\bR_{m'}-\bS_m|^2 \simeq (R_{m'}-S_m)^2 + R_c^2(\Phi_{m'}-\Psi_m)^2,
\label{diffRpSvec}
\eeq
where we used that to leading order $R_{m'}\simeq S_m\simeq R_c$.
In order to deal with the first
term on the RHS of (\ref{diffRpSvec}) we consider the triangle inequality 
\beq
|R_{m'}-S_m| \leq |R_{m'}-R_m|+|R_m-S_m|.
\label{triangleineq}
\eeq
Since $R_m=S_m\cos\gamma_m$, we have 
\beq
|R_m-S_m|\simeq\tfrac{1}{2} R_c\gamma_m^2 ={\cal O}(n^{-\half}).
\label{diffRS}
\eeq
Furthermore, from (\ref{recr}) 
it follows that
\bea
r_{m'}-r_m &\simeq& 2\pi n^{-1} \sum_{\ell=m+1}^{m'}u_\ell \nonumber\\
&\simeq& 2\pi n^{-1}|m'-m|\, u_m\,, 
\label{diffrpr}
\eea
where in the last step we used that for $\ell$ in an interval of length 
$|m'-m| \sim n^\half$ the value of
$u_\ell$ varies negligibly with respect to
its typical value. Hence (\ref{diffrpr}) leads to 
$|r_{m'}-r_m|={\cal O}(n^{-\half})$, which together with (\ref{relRmrm}) 
gives
\beq
|R_{m'}-R_m|=R_{c}^2 n^{-\half} |r_{m'}-r_m|={\cal O}(n^{-\half}).
\label{diffRpR}
\eeq
Upon using (\ref{diffRS}) and (\ref{diffRpR}) in (\ref{triangleineq})
we find that $|R_{m'}-S_m| \leq {\cal O}(n^{-\half})$.
When substituting this in (\ref{diffRpSvec}) and using 
(\ref{diffPhipPsi}) one obtains
that $|\bR_{m'}-\bS_m| = {\cal O}(n^{-\half})$.
This says that any vertex on the polygon through the $S_m$
is, typically, within a distance 
$\sim n^{-\half}$ of a vertex of the polygon through the $R_m$.
In the limit $n\to\infty$
we may therefore indifferently refer to either of them as the perimeter.


\section{Lewis' law}
\label{secLewis}

The average area $A$ of a Voronoi cell is
equal to $A=\lambda^{-1}$, where, here and henceforth, we restore 
the particle density $\lambda$. 
Its average number of sides $\la n\ra$ is topologically constrained to
$\la n \ra = 6$.
One may now investigate the average area $A_n$ of an $n$-sided Voronoi cell.
Lewis' empirical law \cite{Lewis} states that 
$A_n$ is linear in $n$. The law is
often expressed with a single free parameter $c$,
\beq
A_n = [c(n-6)+1]\lambda^{-1},
\label{Lewislaw1param}
\eeq
so as to imply that
$A_6=\lambda^{-1}$, {\it i.e.,} the average area of a cell
having $\la n \ra$ sides is equal to the average area of all
cells; this appealing identity is not, however, a mathematical necessity.
The law (\ref{Lewislaw1param}) 
accounts very well for  data on cell areas coming from a wide
range of different experimental situations \cite{Okabeetal00},
despite deviations often found for the
small-$n$ values $n=3$ and $n=4$.
Lewis' law also seems to apply 
\cite{LeCaerHo90,Flyvbjerg93,SMKS93,Fortes95,FortesTeixeira01} 
to models different from Poisson-Voronoi diagrams, 
whether dynamical or not.  

The numerical estimates of the coefficient $c$ presented in
reference \cite{Okabeetal00}
are in the range $0.199-0.257$. 
In particular, Lauritsen {\it et al.} \cite{LMH93} 
estimate $c=0.23$; 
Drouffe and Itzykson \cite{DI84}, on the basis of their Monte Carlo
simulations for large  $n$, 
observe that $c$ is `very close' to $\frac{1}{4}$;
and Miles and Maillardet \cite{MilesMaillardet82}, on the basis of a
comparison of the cell area with the area of its
fundamental domain \cite{footnote1},
conjecture that $c$ `does not differ greatly' from $\frac{1}{4}$.
It has been noticed \cite{RivierLissowski82} 
that when the `natural' boundary condition $A_2=0$
is imposed on (\ref{Lewislaw1param}), one finds by coincidence
$c=\frac{1}{4}$ exactly, whence
\beq
A_n = \tfrac{1}{4}(n-2)\lambda^{-1}.
\label{Lewislaw0param}
\eeq
Various `derivations' 
of Lewis' law (\ref{Lewislaw1param})
have been proposed \cite{RivierLissowski82,Rivier85,EdwardsPithia94}, 
mostly involving the maximization of
an entropy defined as a functional
of the probability distribution $\{p_j\,|\,j=3,4,\ldots\}$. 
None of these derivations, however, is based on first principles, and
Chiu \cite{Chiu95,Chiureview95} has shown that entropy arguments of this kind
do not allow firm conclusions. Moreover,
numerical data on $A_3$ and $A_4$ 
for Poisson-Voronoi diagrams clearly show 
deviations from linearity and hence rule out
laws such as (\ref{Lewislaw1param}) and (\ref{Lewislaw0param})
for the full (nonasymptotic) regime. 

It suffices to extend the arguments of 
the preceding sections only very slightly
to arrive at a new analytic result.
Let us look at the midpoint distances $R_m$.
By combining equations (\ref{defrhom}), (\ref{deftau}), and (\ref{defrm})
we find that $R_m=R_{\rm av}[1+{\cal O}(n^{-\half})]$.
Furthermore the discussion of subsection \ref{secintonR} has shown that
$R_{\rm av}=R_c[1+{\cal O}(n^{-\half})]$.
It follows that $R_m=R_c[1+{\cal O}(n^{-\half})]$ for all $m$,
where we recall that $R_c=(n/4\pi\lambda)^{\half}$.
Hence the cell perimeter stays within a distance of order unity
from a circle of radius $\sim n^{\half}$. 
Therefore the area
of the $n$-sided 
Voronoi cell is sharply peaked around an average $A_n$ 
for which
we find the asymptotic relation
\beq
A_n \simeq \pi R_{\rm c}^2 = \frac{n}{4\lambda}.
\label{Lewislawaspt}
\eeq
Hence we have demonstrated from first principles that Lewis' law holds to
leading order in an asymptotic expansion for large $n$, and that its
coefficient is $c=\frac{1}{4}$.


\section{Comments and conclusion}

\label{secconclusion}

By setting out to find the probability $p_n$ for
a Voronoi cell to have $n$ sides, we ended up unraveling
the detailed statistical structure of the many-sided Voronoi cell.
We have obtained the asymptotic 
expansion of $\log p_n$ 
up to terms that vanish as $n\to\infty$,
and have shown that a perturbation series may be set up 
for the higher order terms.
We found that the problem is characterized by
two spatial scales which in the limit
$n\to\infty$, and to the order that we calculated, 
become completely disjoint. The first one is the microscopic
scale and the second one the scale of the system size.
\vspace{2mm}

The exact methods and results of this work suggest, in retrospect, 
several heuristic considerations. These may be expected to 
lend to our arguments a flexibility
that will allow them to be applied in more general situations.
We reserve the heuristics, 
together with further extensions of the present results,
to a companion paper \cite{HJHpartII05}.
Our work also opens the way to efficient Monte Carlo simulations
of large Voronoi cells; we also leave this and other numerical
applications for future work.
\vspace{2mm}

Voronoi cells in spatial dimensions higher than two 
have likewise received a great deal of attention in the literature
\cite{Okabeetal00}.
We do not see how to calculate an analytic result analogous to
our equations (\ref{resultpn})-(\ref{resultC}) in dimensions above two.
The reason is that the idea of constructing a Markov process along the
cell perimeter, which was at the root of the present work, 
has no obvious extension to higher dimensions.
Nevertheless, we believe that it is possible
without detailed calculation to transpose
many of the scaling relations found in this work
to higher dimensions.
\vspace{2mm}

The problem of random points in a plane considered here
(the Poisson point
process) is twin to that of random lines in a plane: 
the {\it Poisson line process}. 
Instead of Voronoi cells, random lines define
{\it Crofton cells}, about whose statistics which one may
ask similar questions.
In recent mathematical literature Voronoi and Crofton
cells have often been discussed together
\cite{Calka03b,Hugetal04,CalkaSchreiber05}.
In physics random lines in a plane were
was introduced by Goudsmit \cite{Goudsmit45},
who was interested in the statistics of tracks in a cloud-chamber.
He asked, specifically, about the probability that three or more independent
tracks intersect at almost the same point and converted this question into
one about the area of the Crofton cell.
We are confident that the methods of this paper can be made to work
for the description of the large $n$-sided Crofton cell.
\vspace{2mm}

In metallography large cells were observed by Aboav \cite{Aboav70} in
the arrangement of grains in a polycrystal. 
Aboav's empirical law (also called the Aboav-Weaire law) 
states that the neighbor of an $n$-sided cell
has itself on average \,$m_n=a_0+a_1 n^{-1}$\, neighbors,
where $a_0$ and $a_1$ are constants.
This law has been widely applied to most of the experimental and theoretical
systems discussed in the introduction. 
The law is nevertheless violated, although only slightly,
by numerical results on Voronoi diagrams \cite{LeCaerHo90,Okabeetal00}.
On the basis of the analysis of the present work
we will in a future study \cite{HJHaboav05}
find the true expression for $m_n$ 
when $n$ is asymptotically large.
\vspace{2mm}


\section*{Acknowledgments}
The author thanks Pierre Calka for a discussion and for pointing out
some important literature on Voronoi tessellations.
He thanks Bernard Jancovici for discussions on Coulomb systems.
This work has benefitted from a six month sabbatical period (CRCT) 
awarded to the author by the French Ministry of Education.
The Laboratoire de Physique Th\'eorique of the Universit\'e de Paris-Sud
is associated with the Centre National de la Recherche Scientifique as
research unit UMR 8627.



\begin{thebibliography}{10}
\bibitem{footnote1}
This region, when enlarged by a factor of two in both directions, 
is sometimes referred to as the `fundamental domain'
of the Voronoi cell. It is then the region that should be free of
point particles.
\bibitem{footnote2}
The definition of $T_m$ in reference
\cite{HJHletter05} contains an extra factor
$\cos\gamma_m/\cos\beta_m$. Only $\Pi_m T_m$ intervenes in the
calculation and the condition $G=0$ guarantees that the product of the
extra factors is unity.
\bibitem{footnote3}
In the special case of a regular $n$-sided polygon, $R_{\rm av}$ is the
radius of its inscribed circle and 
${\cal A}-\pi R_{\rm av}^2$ tends to zero as $\sim n^{-2}$
when $n\to\infty$.
However, in the general case we cannot expect a similar decay with $n$.
\bibitem{footnoteimplications} The strong implications of this fact
will be pursued further in reference \cite{HJHpartII05}.
\bibitem{footnote4}
In reference \cite{HJHletter05} our method was presented more compactly 
with an integral only on the single variable $s$.
\bibitem{footnote5}
The term ${\mathbb V}_1$ was called 
${\mathbb V}_0$ in reference \cite{HJHletter05}.
\bibitem{footnotecorr}
Provided that 
$\delta\eta_m$ is sufficiently weakly correlated with the expressions
that multiply it: $\gamma_m^2+\beta_{m+1}^2$, $\tau_m$, and $\tau_m^2$,
respectively, in the discussion of section \ref{secfactor1pV}. 
The weakness of these correlations may be
shown within the present theory; however, we will not elaborate on this point.
\bibitem{footnote6}
In reference \cite{HJHletter05} at the end of section 3 
it should say `inverse temperature' instead of `temperature.'
\bibitem{footnoteangle}
It is easy to verify that,
up to corrections that vanish as $\sim n^{-\half}$, the angle $\phi$ is
equal to the polar angles $\Phi(\phi)\equiv\Phi_m$ and
$\Psi(\phi)\equiv\Psi_m$.
\end{thebibliography}
\end{document}